\documentclass[10pt,preprint]{aastex}
\usepackage{epsfig}
\usepackage{changebar}
\usepackage{emulateapj5}

\bibpunct{(}{)}{;}{a}{}{,}

\slugcomment{To appear in ApJ}

\shorttitle{{\sc Acbar} Power Spectrum}
\shortauthors{Kuo et al.}

\usepackage{amsmath}
\usepackage{amssymb}

\begin{document}

\title{High Resolution Observations of the CMB Power Spectrum with {\sc Acbar}}


\author{C.L. Kuo\altaffilmark{1,2}, 
P.A.R. Ade\altaffilmark{3},
J.J. Bock\altaffilmark{4}, 
C. Cantalupo\altaffilmark{5},
M.D. Daub \altaffilmark{1}, 
J. Goldstein\altaffilmark{6,7}, 
W.L. Holzapfel \altaffilmark{1}, 
A.E. Lange\altaffilmark{8}, 
M. Lueker\altaffilmark{1},
M. Newcomb\altaffilmark{1},
J.B. Peterson\altaffilmark{9}, 
J. Ruhl\altaffilmark{6}, 
M.C. Runyan\altaffilmark{8}, 
E. Torbet\altaffilmark{7}} 

\altaffiltext{1}{Department of Physics, University of California at Berkeley, Berkeley, CA 94720}
\altaffiltext{2}{Department of Astronomy, University of California at Berkeley, Berkeley, CA 94720}
\altaffiltext{3}{Department of Physics and Astronomy, Cardiff University, CF24 3YB Wales, UK}
\altaffiltext{4}{Jet Propulsion Laboratory, Pasadena, CA 91125}
\altaffiltext{5}{Lawerence Berkeley National Laboratory, Berkeley, CA 94720}
\altaffiltext{6}{Department of Physics, Case Western Reserve University, Cleveland, OH 44106}
\altaffiltext{7}{Department of Physics, University of California, Santa Barbara, CA 93106}
\altaffiltext{8}{Department of Physics, Math, and Astronomy, California Institute of Technology, Pasadena, CA 91125}
\altaffiltext{9}{Department of Physics, Carnegie Mellon University, Pittsburgh, PA 15213}


\begin{abstract}

We report the first measurements of anisotropy in the cosmic microwave
background (CMB) radiation with the Arcminute Cosmology Bolometer Array 
Receiver ({\sc Acbar}).
The instrument was installed on the $2.1\,$m Viper telescope at the South Pole 
in January 2001; the data presented here are the product of observations 
up to and including July 2002.
The two deep fields presented here, have had offsets removed by subtracting lead 
and trail observations and cover approximately $24\,{\rm deg}^2$ of sky selected 
for low dust contrast.
These results represent the highest signal to noise observations
of CMB anisotropy to date; in the deepest $150\,$GHz band map, 
we reached an RMS of $\sim8.0\,\mu$K per $5^{\prime}$ beam. 
The 3 degree extent of the maps, and small beamsize of the experiment 
allow the measurement of the CMB anisotropy power spectrum over the range
$\ell = 150-3000$ with resolution of $\Delta \ell=150$.
The contributions of galactic dust and radio sources to the observed 
anisotropy are negligible and are removed in the analysis. 
The resulting power spectrum is found to be consistent with 
the primary anisotropy expected in a concordance $\Lambda$CDM Universe.  

\end{abstract}


\section{Introduction}\label{sec:intro}

Observations of the Cosmic Microwave Background (CMB) provide an 
unique probe of the Universe at the epoch of matter and radiation 
decoupling.
The physics of the early Universe can be described in terms 
of models that yield precise predictions for cosmological
observables.
Within the context of these models, observations the CMB anisotropy 
power can be used to place constraints on the values 
of key cosmological parameters \citep{white94,hu96a}

On sub-horizon size scales, gravity driven acoustic oscillations in 
the primordial plasma give rise to a series of harmonic peaks in 
the CMB angular power spectrum 
\citep{sunyaev70,bond87,hu97b}.
A number of experiments have characterized the power spectrum up to
and including the third acoustic peak 
\citep{lee01,netterfield02,halverson02}. 
These observations have been used to produce constraints 
on cosmological parameters
such as the total energy density and baryon density with
a precision of $\sim 5-10\%$ \citep{lange01,jaffe01,pryke02,abroe02}.

On fine angular scales, the CMB power spectrum is 
exponentially damped due to photon diffusion and the finite 
thickness of the surface of last scattering \citep{silk68,hu97}.
Observations of the CMB power spectrum in this region can be used to
produce independent constraints on cosmological parameters.
For example, the scale of the damping can be used to  
constrain $\Omega_M$ and $\Omega_B$ \citep{white01}.
Recent observations with the Cosmic Background Imager (CBI) have 
provided a first look at the damping tail of the CMB and found it to be 
consistent with models motivated by observations on larger angular 
scales \citep{sievers02}.
On angular scales of a few arcminutes, deep pointings with the CBI 
detect power in excess of that expected 
from primary CMB anisotropy \citep{mason02}. 
In a companion paper to that work, this signal is interpreted as being due to 
the Sunyaev-Zel'dovich Effect (SZE) in distant clusters of galaxies 
\citep{bond02}.
Alternative interpretations have been proposed that explain the excess 
power as being due to local structure and non-standard inflationary models 
\citep{cooray02,griffiths02}.
If the signal is due to the SZE, it's precise characterization would provide 
information about the growth of 
cluster scale structures \citep{holder01,zhang02,komatsu02}.

The Arcminute Cosmology Bolometer Array Receiver ({\sc Acbar})
is an instrument designed to produce detailed images of the CMB in three
millimeter-wavelength bands.
This paper is the first in a series reporting results from the
{\sc Acbar} experiment, and describes the CMB angular power spectrum 
determined from the $150\,$GHz data.
In a companion paper, the {\sc Acbar} CMB power spectrum and results from 
other experiments are used to place constraints on cosmological
parameters \citep{goldstein02}. 
Further papers describing the {\sc Acbar} instrument \citep{runyan02a}, 
SZE cluster searching \citep{runyan02b}, and pointed SZE observations 
\citep{gomez02} are in preparation.

We present a brief overview of the telescope, receiver, and site in
\S~\ref{sec:instrument}. 
Our observing technique, including data editing and calibration are
described in \S~\ref{sec:observations}. 
The analysis of the data is  presented in \S~\ref{sec:analysis}, 
including several developments in the treatment of high sensitivity
ground based CMB observations.
The main results of the paper are presented in \S~\ref{sec:results}.
In \S~\ref{sec:foregrounds}, potential sources of foreground emission
and their treatment in the analysis are discussed.
Tests for systematic error in the power spectrum are discussed in
\S~\ref{sec:systematic}.
In section \S~\ref{sec:conclusion}, we present our conclusions.

\section{{\sc Acbar} Instrument}\label{sec:instrument}

The Arcminute Cosmology Bolometer Array Receiver ({\sc Acbar}) is
a 16 element $235\,$mK bolometer array that has been used to image
the sky in three millimeter-wavelength bands. 
The instrument was designed to make use of
the Viper telescope at the South Pole to produce multi-frequency maps
of the CMB with high sensitivity and high angular 
resolution ($\sim 4-5^{\prime}$). 

The Viper telescope is a $2.1\,$m off-axis aplanatic Gregorian
telescope designed specifically for observations of CMB anisotropy. 
A servo controlled chopping tertiary mirror is used to modulate 
the optical signal reaching the {\sc Acbar} receiver.
The secondary mirror produces an image of the primary mirror at the
tertiary and therefore the motion of the chopping mirror produces minimal 
changes in the illumination pattern on the primary. 
With this system, it is possible to modulate the 16 {\sc Acbar} beams 
$3^{\circ}$ in azimuth in a fraction of a second without 
introducing excessive modulated telescope emission or vibration. 
The primary is surrounded by an additional $0.5\,$m skirt that reflects 
primary spillover to the sky.
To minimize possible 
ground pickup, a reflective conical ground shield completely surrounds the 
telescope, blocking emission from elevations below $\sim 20^\circ$.  

Between 1998-1999, Viper was equipped with the single-element HEMT-based 
CORONA receiver operating at $40\,$GHz. 
These observations produced a detection of the first acoustic peak in the CMB power
spectrum \citep{peterson00}. The {\sc Acbar} instrument was designed specifically
to take full advantage of the unique capabilities of the Viper telescope 
and was deployed to the South Pole in December 2000.  

{\sc Acbar} makes use of microlithographed ``spider-web'' bolometers developed 
at JPL as prototypes for the Planck satellite mission. The detectors are 
cooled by a three stage closed-cycle $^4$He-$^3$He sorption refrigerator
to $235\,$mK, at which point they are background limited.  The fridge and focal 
plane are mounted on the $4\,$K cold plate of a liquid helium and nitrogen 
cryostat.  
The signals from the bolometers are amplified and sampled with 
a 16 bit A/D at a frequency of $2.4\,$kHz

A set of sixteen corrugated feed horns couples the radiation from the 
telescope to the detectors.
The horns are designed to produce nearly Gaussian beams on the sky that
are all of approximately
equal size at all observing frequencies.  The focal plane array projects 
onto a $4\times 4$ grid on the sky with a spacing of $\sim15^\prime$
between array elements.
Behind each beam defining horn there is a filter stack that 
defines the passbands and blocks high frequency leaks.  The band centers 
are 150, 220, and $280\,$GHz with associated bandwidths of 30, 30, and 
$50\,$GHz, respectively.
In the 2002 observations, each of the 8 $150\,$GHz channels typically achieved a 
noise equivalent CMB temperature of $\sim 340\,\mu{\rm K} \sqrt{s}$.
Details of the instrument construction and performance are presented 
by \citet{runyan02a}.

The South Pole station is located at an altitude of $\sim 2900\,$m and
experiences ambient temperatures ranging from $-30^\circ$C to
$-80^\circ$C. In the best winter weather, the perceptible water vapor has
been measured to be $\sim 0.2\,$mm \citep{chamberlin01}. The high
altitude, dry air, and lack of diurnal variations result in a transparent
and extremely stable atmosphere \citep{lay00,peterson02}.  
During the winters of 2001 and 2002, the typical
atmospheric optical depth for the {\sc Acbar} $150\,$GHz channels was measured 
to be
approximately $3\%$.  The entire southern celestial hemisphere is
available year round allowing very deep integrations.
The combination of these unique features and the established infrastructure
for research make the South Pole a nearly ideal site for ground based 
observations of the CMB. 

During an observation, the telescope tracks the position of the observed field.
Due to the proximity of the telescope to the geographic South Pole,
the telescope needs to rotate in azimuth with only small changes
in elevation.
The beams of the array follow a constant velocity $3^\circ$ 
triangle wave in azimuth with a speed fast enough 
that the atmosphere is essentially stationary, and slow enough that the beams
take several time constants to move across a point source on the sky.
Subject to these constraints, the chop frequency was chosen to
be $0.7\,$Hz in 2001 and $0.3\,$Hz in 2002.

The combination of large chop ($\sim 3^\circ$) and small beam sizes
($\sim 4-5'$) make {\sc Acbar} sensitive to a wide range of angular
scales ($150<\ell<3000$), with high $\ell$-space resolution ($\Delta
\ell\sim 150$). Another unique feature of {\sc Acbar} is its multi-frequency
coverage, which has the potential to discriminate between
sources of signal and foreground confusion.
The CMB power spectrum we present in this work is derived from the
$150\,$GHz channel data. In 2001, {\sc Acbar} was configured with four 150 GHz
detectors; this number was increased to eight for the 2002 observations.  
An analysis of the $220\,$GHz and $280\,$GHz data is underway.

\section{Observations}\label{sec:observations}

To minimize possible pickup from the modulation of telescope sidelobes
on the ground shield, 
we restrict the CMB observations to fields with ${\rm EL} \gtrsim 45^\circ$. 
Fortunately, the lowest dust contrast region of the southern sky
is centered at an elevation of EL$\sim 55^{\circ}$ when observed from
the South Pole. 
Several low dust, high declination fields were selected for CMB
observations. 
In general, the {\sc Acbar} observations have focused on producing high
signal to noise maps rather than covering more sky in order to
minimize the sensitivity of the
power spectrum to the details of the noise estimate.
The power spectrum reported in this paper is derived from observations of
two separate fields, which we call {\bf CMB2} and {\bf CMB5}. 
Each field was chosen to include a bright quasar,
PMN J0455-4616 ($\alpha_{J2000}=4^h55^m50.8^s$, $\delta_{J2000}= 
-46^\circ 15^\prime 59^{\prime\prime}$) and PMN J0253-5441 
($\alpha_{J2000}=2^h53^m29.2^s$, $\delta_{J2000}=
-54^\circ 41^\prime 51^{\prime\prime}$), respectively. 
As described in \S\ref{sec:pointing}, the pointing model is derived from
frequent observations of quasars, Galactic HII regions and planets.
The images of the guiding quasars produced during the CMB observations
provide stringent, independent 
constraints on beam sizes and pointing accuracy. 
The two fields are sufficiently separated to be considered independent, 
and the signal correlation between them can be safely ignored.

\begin{figure*}[t]
\centerline{
\psfig{figure=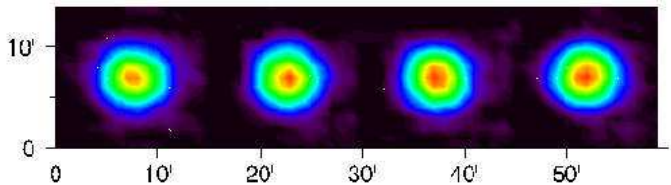,width=4in,height=1in,angle=0}}
\caption{The beams of the $150\,$GHz array elements determined
from observations of the CMB2 guiding quasar in the 2001 season.
These images represent an
average over the entire observation period and include any
distortions due to changes in pointing or beamsize.}
\label{fig:cmb2_qmap}
\end{figure*}

\begin{figure*}[t]
\centerline{
\psfig{figure=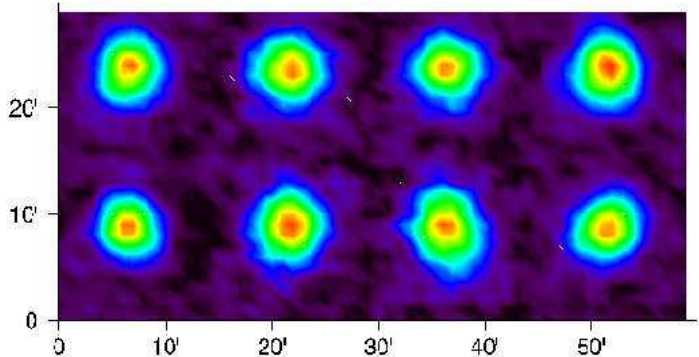,width=4in,height=2in,angle=0}}
\caption{The beams of the $150\,$GHz array elements determined
from observations of the guiding quasar in the CMB5 field.
These images represent an
average over the entire observation period and include any
distortions due to changes in pointing or beamsize.}
\label{fig:cmb5_qmap}
\end{figure*}

\subsection{The Lead-Main-Trail Scan Strategy}\label{subsec:lmt}

During the CMB observations, the telescope tracks a position on the sky while
the chopper sweeps the beams back and forth three degrees in azimuth.
The motion of the chopper introduces systematic offsets into the data 
that must be treated in the analysis.
As will be described in \S\ref{sec:pointing}, the motion of the  
chopper causes the beams to move slightly ($\sim 2^{\prime}$) in 
elevation as they sweep in azimuth. 
This motion modulates the atmospheric emission and introduces an 
$\sim 5\,{\rm mK} (T_{air}/200\,{\rm K})(\tau/0.03)$ signal at an 
elevation of $45^\circ$, where $\tau$ is the zenith optical depth
at $150\,$GHz. 
Although the telescope optics were designed to produce a stationary
illumination pattern on the primary mirror, the illumination on the 
secondary mirror changes substantially with chopper angle. Snow 
accumulation or temperature differences on the secondary mirror will be 
modulated and contribute a second source systematic signal that varies
with the chopper position.

The effects described above are collectively referred to as the 
{\it chopper synchronous offset}. Changes in the effective
temperature of the atmosphere and optics throughout
an observation can result in a time variation of this signal.
In order to minimize the chopper synchronous offset, each field 
is observed in a rapid Lead-Main-Trail ({\it LMT}) sequence where the 
Main field is led and trailed by two identical Lead 
and Trail observations each with half the integration time of the 
main observation.
The CMB power spectrum is derived
from the differenced map, Main-(Lead+Trail)/2. If the LMT 
switching is performed faster than the time scale of any drift in the
chopper synchronous offset, then the offset can be eliminated.
Compared to the differencing of two equally weighted fields, the LMT 
observation further 
removes a linear drift in any chopper synchronous offset.
In practice, the Lead field is tracked for 30 seconds; the telescope
is moved $+3^\circ\sec\delta$ (in increasing RA) to track the 
Main field for 60 seconds; the telescope is moved another 
$+3^\circ\sec\delta$ to the Trail field for 30 seconds of integration.
After completing the LMT cycle, the elevation of the telescope
is decreased by $1^{\prime}$ and the process is repeated $\sim100$ times 
to build up a continuous 2-dimensional map.  
The CMB2 and CMB5 fields actually consist of two largely overlapping sub-fields
offset by $0.5^{\circ}{\rm sec}(\delta)$ in RA. 
The total observation time for each field is split roughly equally between the 
subfields which are referred to as CMB2a,b and CMB5a,b.

\subsection{Data Cuts}\label{subsec:data}

In order to minimize the potential contamination of the data by systematic
errors, we introduced a number of conservative data cuts.
All 4 hour observations containing refrigerator temperatures higher 
than $250\,$mK cut from the data set due to their uncertain calibration. 
Cryogenics fills, refrigerator cycles, computer and telescope maintenance,
and bad weather all limit observation time. 
Including galactic observations and calibrations, the effective observing 
efficiency was $\sim 60\%$ for the winter of 2002. 
The effects of cosmic ray interactions with the detectors were removed 
by discarding any chopper 
sweep containing a point varying from the mean of the raw $2.4\,$kHz data 
samples by $>6 \sigma$.
Due to the low filling factor of the micro-mesh detectors, this results 
in an insignificant loss of data. 
Before performing the LMT differencing, the data are examined for the presence
of excessive chopper synchronous offsets that indicate the accumulation 
of snow on the mirrors.
The offset level is bimodal between the two extremes and the amount of data 
cut depends only weakly on the cut level. 
Files with snow accumulation are discarded, resulting in a loss of approximately
$40\%$ of the remaining data. 
The details of the cut level and amount of discarded data have no significant 
effect on the resulting maps or power spectrum. 
After implementing the data cuts described above, we retained $\sim 400$ 
hours of observation on CMB2 during July--August of 2001 and April of 2002. 
For CMB5 we retained $\sim 800$ hours of data gathered during 
April--July of 2002.
As described in \S\ref{subsubsec:nwcm}, we cut an additional 
$\sim 25\%$ of the remaining data that we determined to be corrupted by 
atmosphere induced correlations between array elements and adjacent scans.

\subsection{Pointing}\label{sec:pointing}

The Viper telescope pointing model is developed from observations
of bright galactic sources, planets, and quasars.
Daily observations of the compact galactic sources RCW38, RCW57 and MAT6a 
are used to monitor pointing offsets and check the pointing model.
The pointing model contains terms for encoder offsets, collimation error, telescope 
flexure, and tilt of the AZ ring. 
When the telescope is pointed at the equator, the detector beams sweep out a strip 
of approximately constant elevation under the motion of the chopper.
However, at the elevation at which the CMB observations are made, the beams sweep 
out arcs with amplitude of approximately $2\,^{\prime}$ across the $3^{\circ}$ chop. 
Observations of planets and galactic sources at a number of positions in the 
chopper throw have been used to characterize this effect. 
For each array element, we determine a pointing model that is a function of
the elevation, azimuth, and chopper encoders.
After applying the model, we find the RMS pointing error for the 2001 season
to be $\sim 1.4^{\prime}$ in RA and $\sim 30^{\prime\prime}$ in declination.
The majority of the spread in 2001 was due to a malfunctioning sensor of 
the chopper angle.
After this sensor was replaced in January 2002, the pointing RMS was determined 
to be $\sim 20^{\prime\prime}$ in RA and $\sim 20^{\prime\prime}$ in declination.
Due to early concerns about the accuracy of the pointing, we chose each CMB field
to have a bright ( $\gtrsim 1\,$Jy at $5\,$GHz)
quasar in the middle.
In this way, we have an continuous monitor and independent check of the accuracy
of the pointing.

\subsection{Beams}\label{subsec:beams}

High signal-to-noise maps of planets (Mars and Venus in July 2001,
and Venus in September 2002) are used to accurately measure the beam 
patterns of the array elements. 
The measured maps were deconvolved with a constant temperature circular disk 
corresponding to the planet sizes (typically small compared to the beams), to 
determine the instantaneous beam parameters.

The coadded image of the guide quasar is used to determine the effective 
beamsizes at the center of the map and includes any smearing due
to drifts in pointing that occur over the period in which the data are acquired.
Focal plane maps for the array in both the 2001 and 2002 configurations
are shown in Figures~\ref{fig:cmb2_qmap} and \ref{fig:cmb5_qmap}.
In Table~\ref{tab:qbeams}, we list the FWHM beam diameters derived from
Gaussian fits to the quasar images for both years. These beam parameters were
used in the analysis of CMB power spectrum.
These results are consistent with the beamsizes measured on planets and 
the observed pointing RMS determined from repeated observations of galactic 
sources.
Some of the beams in CMB5 quasar image appear to be slightly elongated in 
declination; we believe this to be due to the accumulation of frost on the 
dewar window during the 2002 observations. We estimate the uncertainty in
the FWHM of the beams to be less than $3\%$.

\begin{table*}
\caption{\label{tab:qbeams}150 GHz Beam Sizes}
\small
\begin{center}
\begin{tabular}{ccccc}
\hline\hline
\rule[-2mm]{0mm}{6mm}
year & channel & ${\rm FWHM}_1(^\prime)$ &${\rm FWHM}_2(^\prime)$
& {\rm P.A.}\tablenotemark{a} (degree)\\
\hline
2001 & {\rm C1}   & 5.46       & 4.76      & 178\\
2001 & {\rm C2}   & 5.10       & 4.81      & 176\\
2001 & {\rm C3}   & 5.10       & 4.76      & 161\\
2001 & {\rm C4}   & 5.19       & 4.68      & 3\\
\\
2002 & {\rm D3}   & 5.71       & 4.90      & 68\\
2002 & {\rm D4}   & 5.50       & 5.45      & 121\\
2002 & {\rm D6}   & 5.53       & 5.25      & 7\\
2002 & {\rm D5}   & 5.69       & 5.15      & 87\\
2002 & {\rm B6}   & 4.94       & 4.40      & 111\\
2002 & {\rm B3}   & 5.66       & 5.20      & 69\\
2002 & {\rm B2}   & 6.79       & 5.34      & 105\\
2002 & {\rm B1}   & 5.46       & 4.67      & 43\\

\hline
\end{tabular}
\end{center}
$\;\;\;\;\;\;\;\;\;\;\;\;\;\;\;\;\;\;\;\;\;\;\;\;\;
\;\;\;\;\;\;\;\;\;\;\;\;\;\;\;\;\;\;\;\;\;\;\;\;\;
\;\;\;\;\;\;\;\;\;\;\;\;\;\;^a$ 
Position angle from +RA, counterclockwise.
\normalsize
\end{table*}

Due to aberrations in the telescope optics, the beam sizes increase slightly
near the extremes of the chopper travel.
This effect was characterized by mapping planets and galactic sources
over a range of pointings offset in RA.
The telescope beams were simulated with a ray-tracing package and 
the resulting aberrations were found to be consistent with the observations.
The FWHM of the beams typically changes by $\sim 5\%$ over the entire chopper 
throw, and the throughputs are found to be conserved.
The measured beams are found to be very close to symmetric Gaussians and we 
describe their shapes in terms of their small distortions. 
We parameterize the aberrations with three parameters: the distortions 
along RA, DEC, and an axis inclined at a 45$^\circ$ position angle. 
When the distortions are small, all three parameters are small and can be 
treated with a perturbation analysis. In sections \S\ref{subsec:matrix} and 
Appendix~\ref{app:theorymatrix}, we describe how changes in the beam shape are 
treated in the computation of the power spectrum.

\subsection{Calibration}\label{subsec:cal}

The calibration of {\sc Acbar} is described in detail by \cite{runyan02a}; 
here we will present a brief summary.
The planets Venus and Mars serve as the primary calibrators for the {\sc Acbar} 
CMB observations.
Due to the lack of a significant atmosphere and dielectric emission by the 
soil, the spectrum of Mars closely approximates that of a blackbody.
The brightness temperature of Mars varies as a function of its distance
from the sun with mean Heliocentric value of $\sim 206\,$K at $150\,$GHz.
The FLUXES\footnote{http://star-www.rl.ac.uk} software package uses the 
brightness temperature adopted by \citet{griffin86} and an 
accurate ephemeris to compute instantaneous values for the Martian brightness 
temperature. 
Following \citet{griffin93} we assign an overall uncertainty to the 
temperature of Mars of $5\%$.
The brightness temperature of Venus is a function of frequency;
following \citet{weisstein96}, we fit the brightness temperature of
Venus with a linear fit to the published data.
For the {\sc Acbar} $150\,$GHz band, we adopt a brightness temperature of $297\pm{22}\,$K,
close to the value of $294\pm22\,$K reported by \citet{ulich81} for a band centered
about $150\,$GHz. 
From the South Pole, planets are only intermittently available for observation 
and are always at elevation $<30^{\circ}$.
Observing them typically requires lowering a section of the groundshield and 
waiting for the sky to clear from the zenith to the horizon. 
For these reasons, it is impractical to use planets to regularly monitor the 
calibration of the instrument.

The HII region RCW38 is bright, compact, and has a declination similar
to our CMB fields; therefore, it serves as an excellent secondary calibrator. 
The responsivities of the {\sc Acbar} detectors are a function of the 
background loading and change slightly with elevation.
Behind a small hole in the tertiary mirror, we have installed a chopped
thermal load that produces a constant optical signal which is used to monitor the 
detector responsivity as a function of elevation.
The difference in atmospheric attenuation between the planet and RCW38 
observations must also be taken into account. 
Sky dips are used to measure the atmospheric opacity
and correct for the effects of atmospheric attenuation.
Because the emission from RCW38 is somewhat extended ($\sim 3^{\prime}$ 
FWHM) and
beamsizes of {\sc Acbar} differ slightly between the 2001 and 2002 observations, 
the integrated flux (to an angular radius of $8^{\prime}$) is used as our 
secondary calibration standard. 

The integrated galactic source flux is given by
\begin{equation}
S_I = \int I_P(\nu) f(\nu) d\nu \frac{\int V_G d\Omega }{\int V_P d\Omega} 
\left(\frac{R_P}{R_G}\right) 
\left({\frac{e^{-\tau_P\csc(\theta_P)}}{e^{-\tau_G\csc(\theta_G)}}}\right) \,,
\end{equation}
where $R_P$ and $R_G$ are the responsivities of 
of the detectors when observing the planet and galactic source, 
$V_P$ and $V_G$ are the 
measured voltages as a function of angular position, 
$\tau_P$ and $\tau_G$ are the zenith optical depths of the atmosphere, 
$\theta_P$ and $\theta_G$ are the elevations of the planet and galactic 
source observations, $I_P(\nu)$ is intensity of the planet, and $f(\nu)$
is the spectral response of the system as measured in the lab with
a Fourier transform spectrometer as described by \cite{runyan02a}. 
Both the Mars and Venus calibrations produce a consistent integrated flux
for RCW38 at $150\,$GHz of $S_I=145\pm7\,$Jy where the uncertainty 
reflects the scatter among the array elements and measurements.
Including the uncertainty for the flux of the planets, the detector 
responsivities, and integral over solid angle, we find 
$S_I=145\pm13\,$Jy which is dominated by the uncertainty in the planet 
brightness temperature.

We can now use RCW38 to monitor the calibration of the instrument and
transfer the calibration to our CMB observations. Observations of RCW38
are scheduled to bracket each $\sim5$-hour CMB observation. The zenith optical
depth at $350\,\mu$m is measured every 15 minutes with a tipping
radiometer \citep{peterson02} and is extrapolated to $150\,$GHz using the
observed scaling between the measured {\sc Acbar} $150\,$GHz opacity and
$350\,\mu$m tipper optical depths.  After correcting for varying
atmospheric attenuation, the integrated RCW38 flux is found to be
constant over all observations with an RMS scatter of $\lesssim 4\%$.
Because both of the CMB fields are observed at elevations similar to
RCW38, the responsivity can be considered to be constant.  Although small,
the difference in optical depth between the RCW38 calibration and the CMB
field must still be taken into account.  The conversion from observed
bolometer voltage signals to CMB temperature differences is given by
\begin{equation}
\frac{dT_{CMB}}{dV_{bolo}} = \frac{S_I}{\int \frac{dB(\nu)}{dT_{CMB}} 
f(\nu) 
d\nu \int{V_G\, d\Omega}}
\left({\frac{e^{-\tau_G\csc(\theta_G)}}{e^{-\tau_{CMB}\csc(\theta_{CMB})}}} 
\right) \,,
\end{equation}
where $B(\nu)$ is the intensity of a $2.73\,$K CMB blackbody, 
and $\theta_{CMB}$ is the elevation of the CMB observation. 

Given the stability of the frequent RCW38 observations, the statistical
accuracy of the calibration transfer from RCW38 to the CMB fields is quite
good.  The uncertainty in the calibration is a combination of the
uncertainty of the planet brightness temperatures, beamsize measurements,
spectroscopy, and accuracy of the calibration transfer. 
Assuming a common overall uncertainty in the calibration of Mars and Venus,
we estimate the uncertainty in the temperature calibration of the observed 
CMB anisotropy to be $\pm 10\%$.

\section{Analysis}\label{sec:analysis}

Algorithms for the analysis of total power CMB data are well developed and 
have been tested on balloon-borne experiments \citep{netterfield02, lee01}. 
However, the ground-based {\sc Acbar} experiment is subject to constraints 
which require significant departures from the standard analysis
algorithms. In this Section, we outline the {\sc Acbar} analysis and highlight
its unique features. 

\subsection{Maximum Likelihood Map and Noise Weighted 
Coadded Map}\label{subsec:map}

The first step of the analysis is to produce a map from the timestream
data. 
Suppose $d_\alpha\;$($\alpha=1..n_t$)
\footnote{Throughout the paper the Greek indices will be used to 
  represent the timestream sample, and the Latin indices will
  be used to label spatial ``pixels''.} are $n_t$ time-ordered measurements of CMB 
temperature. This vector can be separated into the noise component
$n_\alpha$ and the signal component $\sum_jA_{\alpha j}T^0_{j}$, such that
$d_\alpha=n_\alpha+\sum_jA_{\alpha j}T^0_{j}$, or in matrix representation,
${\bf d}={\bf n}+{\bf AT^0}$. Here $T^0_{j}$ ($j=1..n_p$) are the 
CMB temperatures on each sky pixel $j$, after being convolved with the ``beam'', or 
the experimental response function. 
For these observations, this will be a Lead[-1] Main[+2] Trail[-1] three point 
beam function.
${\bf A}$ is the $n_t\times n_p$ pointing matrix where
 $A_{\alpha j}=1$ if $\alpha\in j$ (measurement $\alpha$ corresponds to sky 
pixel $j$); and $A_{\alpha j}=0$ otherwise. If ${\bf n}\gg {\bf AT^0}$ 
(as is the case for the {\sc Acbar} data), 
the time stream correlation can be determined from the data itself; 
$\langle n_\alpha n_\beta \rangle \equiv N_{\alpha \beta}\sim 
\langle d_{\alpha}d_{\beta}\rangle$.

In general, a {\it map} ${ {\bf T}}$ is produced from a
linear combination of the timestream ${\bf d}$;
$$
{ {\bf T}}\equiv {\bf L d},
$$
where ${\bf L}$ is a $n_p\times n_t$ coefficient matrix. The map ``pixel" space
correlation matrix is given by
$$  
\langle { {\bf T}}\;{ {\bf T}}^t\rangle=
	  {\bf L}\langle {\bf d}{\bf d}^t\rangle {\bf L}^t=
	  {\bf L}{\bf N}{\bf L}^t+{\bf L}{\bf A}\langle
	  {\bf T^0}{\bf T^0}^t\rangle {\bf A}^t{\bf L}^t\\
$$
\begin{equation}
\equiv
	  {\bf C}_N+{{\bf C}}_T.\label{corr}
\end{equation}

\subsubsection{Maximum Likelihood Map}
In balloon-borne or satellite experiments, the matrix ${\bf L}$ is usually 
taken to be
\begin{equation}
  {\bf L}={\bf L_m}
  =({\bf A}^t{\bf N}^{-1}{\bf A})^{-1}{\bf A}^t{\bf N}^{-1},\label{Bm}
\end{equation}
and the resulting map ${ {\bf T}}_m$ is the {\it maximum likelihood map}.
Using eq.~[\ref{corr}], the correlation matrix for 
the maximum likelihood map can be derived: 
\begin{equation}
  \langle { {\bf T}}_m\;{ {\bf T}}_m^t\rangle=
  ({\bf A}^t{\bf N}^{-1}{\bf A})^{-1}+\langle
	  {\bf T^0}{\bf T^0}^t\rangle.
\end{equation}
It was shown by several authors \citep{tegmark97b,ferreira00} that a maximum likelihood 
map is not only a good visual representation of the data, but also
a lossless way to compress CMB information; the power 
spectra derived from the map and directly from the timestream will 
have the same uncertainty.

However, for a ground based experiment like {\sc Acbar}, the 
vertical gradient in atmosphere emission restricts the 
observation to constant elevation strips. Furthermore, the special
location of the Viper telescope ($\sim 1$ km from the geographic 
South Pole) results in maps with little cross-linking in declination.
In addition, the large angular scale structure in the timestream
is dominated by atmospheric fluctuations.
The lack of cross-linking, and the need to remove large angular scale 
($\gtrsim 1^{\circ}$) sky noise results in numerical problems in the
computation of the maximum likelihood map and an alternative
analysis is required. 

\subsubsection{Demodulation Analysis}
One can select certain linear combinations of ${\bf d}$
which are believed to be free from sky noise contamination. 
In this case, the coefficient matrix ${\bf L}$ is formed by rows of 
sky ``modulation'' patterns, which are usually taken to be 
edge-tapered sinusoidal functions. This method is known as 
the {\it demodulation analysis}, and has been used successfully by several 
ground based CMB experiments \citep{netterfield97, miller99, peterson00}. 

One of the disadvantages of demodulation analysis is that the 
choice of modulation patterns is arbitrary, and could result in the loss
of useful information. 
In other words, the choice of ${\bf L}$ could 
result in a power spectrum with uncertainties considerably larger than the 
those found from the maximum likelihood map \citep{tegmark97b}. 
Furthermore, the modulated data, or the ``map'' ${\bf L d}$, has little 
to do with the actual CMB temperature distribution on the sky, making it 
difficult to treat foregrounds and test for systematic errors.
We performed a demodulation analysis of the {\sc Acbar} data presented in 
this paper as an early check of our analysis technique. The resulting
power spectrum was found to be roughly consistent with that presented 
here, although with somewhat larger uncertainties.

\subsubsection{Noise Weighted Coadded Map}\label{subsubsec:nwcm}

Due to time varying atmospheric emission, there are large signals in the 
data timestream.
These signals are easy to remove from the timestream, however, they cannot
be simply removed from the final map. 
Therefore, it is essential that the corrupted modes corresponding to the 
atmosphere be removed before the data are coadded to form a map. 
We propose to use the {\it cleaned noise-weighted coadded map}
as an intermediate analysis step. The operation of the 
``corrupted mode projection'' matrix ${\bf \Pi}$ on the timestream
${\bf d}$ projects out the undesired modes, resulting in the cleaned 
timestream data ${\tilde{\bf d}}\equiv {\bf \Pi d}$. We take ${\bf \Pi}$ to be 
a block-diagonal matrix, where the length of each block corresponds to 
a chopper sweep. Each block removes certain modes in each data 
strip. If ${\bf U}$ consists of $m$ columns of linearly independent undesired 
modes, the projection matrix block can be defined as ${\bf \Pi_b}\equiv 
{\bf I}-{\bf U}({\bf U}^t{\bf U})^{-1}{\bf U}^t$ \citep{tegmark97b}. It is 
easy to see ${\bf \Pi_b}{\bf U}=0$, and ${\bf \Pi_b}^t={\bf \Pi_b}$.

With the corrupted modes removed,
we can define the cleaned noise weighted coadded map as
$$
{\tilde T}_i=\lambda_i \sum_{\alpha\in i}\omega_\alpha \tilde d_\alpha.
$$
$\lambda_i\omega_\alpha$ is the ``weight'' associated with 
measurement $\alpha$ on pixel $i$. The normalization condition
requires that
$$
\lambda_i\sum_{\alpha\in i}\omega_\alpha=1.
$$
The cleaned map can be represented as a linear combination of the raw 
timestream;
\begin{equation}
{\tilde{\bf T}}\equiv  {\tilde{\bf L}}{\bf d},\;\;\; {\rm where}\;\;\;
{\tilde{\bf L}}\equiv {\bf \Lambda}{\bf A}^t {\bf \Omega} {\bf \Pi}.
\label{ourb}
\end{equation}
Here ${\bf \Omega}$ and ${\bf \Lambda}$ are diagonal matrices,
$$
\Omega_{\alpha\beta}=\delta_{\alpha\beta}\omega_{\alpha},\;\;\;\;
\Lambda_{ij}=\delta_{ij}\lambda_{i}.
$$
In practice, each sample $\alpha$ is weighted on its inverse variance
after the projection of the corrupted modes,
$$
\omega_{\alpha}=\frac{1}{\sigma_\alpha^2},
\;\;\;\;\;\sigma_{\alpha}^2=\sum_{\beta\gamma}\Pi_{\alpha\beta}N_{\beta\gamma}
\Pi_{\alpha\gamma}.
$$

To clarify the relation between the maximum likelihood map and the noise
weighted coadded map, we investigate the special case where the noise 
${\bf N}$ is diagonal ($N_{\alpha\beta}=\delta_{\alpha\beta} \sigma_\alpha^2$) and 
no modes are projected out (${\bf \Pi}={\bf I}$), 
\begin{gather*}
{\tilde T}_i=\sum_\alpha {\tilde L}_{i\alpha}d_\alpha=
\frac{\sum_\alpha d_\alpha A_{\alpha i}/\sigma_\alpha^2}
     {\sum_\gamma A_{\gamma i}/\sigma_\gamma^2}\\
     =\left[({\bf A}^t{\bf N}^{-1}{\bf A})^{-1}{\bf A}^t{\bf N}^{-1}{\bf d}
       \right]_i.
\end{gather*}
In this case, ${\tilde {\bf T}}$ is identical to the lossless maximum likelihood 
map. In general, ${\tilde {\bf T}}$ will not be lossless; however, if the off 
diagonal terms of ${\bf N}$ are small, ${\tilde {\bf T}}$ will approach
the lossless map. This is the case for {\sc Acbar}.
     
From eq.~[\ref{Bm}], and $\langle {\bf d}\rangle = \langle {\bf n}\rangle
+{\bf AT^0}={\bf AT^0}$, it is easy to show that the maximum 
likelihood map is an unbiased estimate of the CMB sky temperature: $\langle 
{{\bf T}}_m\rangle={\bf T^0}$. However, this is not the case for
${\tilde {\bf T}}$, since
\begin{equation}
\langle{\tilde {\bf T}}\rangle={\bf K}{\bf T^0}
       \neq {\bf T^0};\;\; {\rm where} \;\; {\bf K}
\equiv{\tilde {\bf L}}{\bf A}.\label{kcorrect}
\end{equation}
Therefore the transformation of the theory matrix 
$ {{\bf C}}_T=
{\bf K}\langle{\bf T^0T^{0t}} \rangle {\bf K}^t$ in eq.~[\ref{corr}] 
is nontrivial and has to be calculated in the coadding process. 
Explicitly, 
\begin{equation}
K_{ij}=\lambda_i\!\!\sum_{\alpha\in i;\,\beta\in j}\!\!\omega_\alpha
\Pi_{\alpha\beta}\;\label{k}
\end{equation}
is the transformation matrix between the sky and our 
``map'' space. Hereafter we drop the tilde and simply write $\tilde {\bf T}$ as 
${\bf T}$. The untransformed sky temperature is denoted by ${\bf T^0}$. 

Note that the contaminated mode projection matrix ${\bf \Pi}$ is completely arbitrary 
and can be made to adapt to the conditions of a given observation. 
We take ${\bf \Pi}$ to be block-diagonal, each block ${\bf \Pi_b}$ removes polynomial 
functions of certain orders in each chopper sweep.
The order of the polynomial to be removed depends on the atmospheric conditions
during the observation. In this way, data corrupted by atmospheric emission on large 
angular scales
can still contribute to the determination of the small scale power.
Quantitatively, the polynomial order is chosen to be the lowest such that the 
residual correlation 
between sweeps in the ``stare'' (data strip of constant DEC pointing) is less than $5\%$.
If after removing a 10th order polynomial a residual correlation greater than
$5\%$ remains, the data are discarded. To avoid bias in noise estimation the order of the 
polynomial to be removed is constant over 20 minutes duration. 
The correction matrix ${\bf K}$ for the map is built by coadding the correction
matrices for the individual scans with the appropriate noise weighting. 
The resulting correction matrix ${\bf K}$ was used in a series of Monte 
Carlo simulations to show the band-powers themselves are unbiased by the adaptive
removal of contaminated modes. 

\subsection{Noise Matrix}\label{subsec:matrix}

The precise determination of the noise correlation properties of the data 
is a non-trivial task for high-sensitivity ground based experiments. 
From eqs.[\ref{corr},\ref{ourb}],
the calculation of noise correlation matrix $C_N$ for ${\bf T}$ 
is straightforward once the timestream noise ${\bf N}$ is known:
\begin{equation}
{C}_{N\{ij\}}=
\lambda_i\lambda_j\!\!\sum_{\alpha\in i;\;\;\beta\in j}\!\!\omega_\alpha
\omega_\beta \sum_{\gamma\delta}
\Pi_{\alpha\gamma}N_{\gamma\delta}\Pi_{\beta\delta}\;,\label{cn}
\end{equation}
or in matrix notation,
$$
{{\bf C}}_N={\bf \Lambda}{\bf A}^t {\bf \Omega} {\bf \Pi}
{\bf N}{\bf \Pi}^t{\bf \Omega}{\bf A}{\bf \Lambda}.
$$
This is just a weighted average of ${\bf \Pi}{\bf N}{\bf \Pi}^t$. 
However, since the noise in the timestream is far from stationary, the 
determination of ${\bf N}$ requires careful treatment 
(see Appendix~\ref{app:noise}).
We are confident that the noise correlation for the data presented 
here has been determined to to better than $5\%$. 

\subsection{Theory Matrix}\label{subsec:theory}

The observed CMB temperature $T'(\theta,\phi)$ is the convolution
of the real temperature $T(\theta,\phi)$ with the experimental response 
function $B(\theta,\phi)$. For a LMT subtracted experiment, $B(\theta,\phi)$ is the 
convolution of the beam $B_0(\theta,\phi)$ with the beam-switching pattern

\begin{equation} 
\Xi({\hat {\bf r}})=\delta({\hat {\bf r}})-[\delta({\hat {\bf r}}-{\hat {\bf r}}_1)+
\delta({\hat {\bf r}}-{\hat {\bf r}}_2)]/2,\label{lmt}
\end{equation}

where $\delta$'s are the Dirac $\delta$- functions. For {\sc Acbar}, 
${\hat {\bf r}}_1=(0,\Delta \phi)$ and ${\hat {\bf r}}_2=(0,-\Delta \phi)$ , 
where $\Delta \phi=3^{\circ}\csc\theta$. The Gaussian temperature anisotropy 
can be described by the angular power spectrum 
$C_{\ell}$, or the mean square of the multipole moments: $C_{\ell}=\langle a_{\ell m}^*a_{\ell m}\rangle\;.$ When the observations are restricted to
small fractions of the celestial sphere, the spherical harmonic series can
be approximated by two dimensional Fourier series. In this limit, 
the 2-D power spectrum density (PSD) is the 
product of $|{\cal B}({\bf k}={\bf \ell}/2\pi)|^2$ and $C_{\ell}$,
where ${\cal B}$ is the Fourier transform of $B$.
\citep{saez_1996}.
For a given power spectrum and observational response function, 
the theory matrix gives the
temperature correlation between pixel pairs in the map. Since the
PSD and the autocorrelation function (ACF) form a Fourier transform pair, the
theory matrix can
be calculated efficiently with Fast Fourier transform (FFT) technique. 

However, the noise weighted coadded map defined in eq.[\ref{ourb}] combines sky 
temperatures measured by channels with different beam sizes at different 
chopper positions. Therefore, each pixel has different beam parameters, and
strictly speaking, the FFT algorithm is not applicable.
The straightforward calculation of the theory matrix and its derivatives 
with respect to band-powers involves a computationally expensive four 
dimensional integral for each pixel pair.
A common way to 
deal with this problem is to average the beam for the entire map, and 
generate the theory matrix from the average beam \citep{wu01}.

We have developed an alternative approach that employs a semi-analytic 
correction formula to treat the variation of beam sizes in the map.
The beams for each measurement are coadded accounting for the position of each 
sample and allowing for changes in the beamshape. 
Different pixel pairs then receive an individual correction to the overall FFT 
result, depending on the pixel-beam parameters.
The full derivation of this technique is given in 
Appendix~\ref{app:theorymatrix}. 
Due to the small changes in beamshape ($\sim 5\%$ FWHM), the effect of the 
beamshape correction is small and does not result in significant changes in 
the measured power spectrum.

\subsection{Power Spectrum Estimation}\label{subsec:power}

The first step in the estimation of the power spectrum is the determination of a 
function describing the likelihood that the data are described by a given model.
Assuming Gaussianity, the likelihood function is 
\begin{equation}
{\cal L}\propto |{\bf C}|^{-1/2}\exp\left[-\frac{1}{2}{\bf T^tC^{-1}T}\right],
\end{equation}
where ${\bf C}={\bf C}_T+{\bf C}_N$. 
Once ${{\bf C}}_N={\tilde{\bf L}}{\bf N}\tilde{\bf L}^t$, ${\bf K}
={\tilde {\bf L}}{\bf A}$, and ${{\bf T}}={\tilde {\bf L}}{\bf d}$
are calculated from the time stream, the anisotropy power spectrum can
be readily estimated by the standard maximum likelihood method. The only
difference is the replacement of ${\bf C}_T$ with 
${\bf K}{\bf C}_T{\bf K}^t$. 

\subsubsection{Signal-to-Noise Eigenmodes And Foreground Removal}

Following \citet{bunn97} and \citet{bond98}, the data are further compressed 
using signal-to-noise eigenmode truncation (Karhunen-Lo\a' eve 
transformation). 
With slight modifications, this can be done simultaneously with the removal
of foreground templates. 
Our foreground removal method is conceptually the same as 
the constraint matrix formalism \citep{bond98, halverson02}, or the pseudo-inversion 
of \citet{tegmark97b}. The idea is to replace the inverse of Hermitian square root 
or Cholesky decomposition (${\bf C}_N^{-1/2}$) in eq.[A4] of 
\citet{bond98} with a non-square ``whitening" matrix. 
This procedure is described in detail in Appendix~\ref{app:sn}. 
The foreground modes removed from the CMB2 and CMB5 fields are 
described in \S~\ref{sec:foregrounds} and listed in 
Tables~\ref{tab:foreground1} and \ref{tab:foreground2}.

\subsubsection{Iterative Quadratic Approach to Maximum Likelihood Band-Power}\label{subsubsec:iterative}

The anisotropy power spectrum is parameterized by the 
band-power ${\bf q}$, where
\begin{equation}
D_\ell\equiv \frac{\ell(\ell+1)}{2\pi}C_\ell=\sum_Bq_B\chi_{B\ell}. \label{qb}
\end{equation}
A convenient choice of $\chi_{B\ell}$ are ``tophat'' 
functions, i.e., $\chi_{B\ell}=1$ for $\ell \in B$; and
$\chi_{B\ell}=0$ for $\ell \not\in B$.
The maximum likelihood ${q_B}$s are estimated iteratively
using the quadratic iteration method \citep{bond98}. 
Near the extrema, the log of the likelihood function can be 
approximated by a multidimensional quadratic function. 
The true maximum of the original $\cal L$ can then be found by an 
iterative process:
\begin{equation}
{q_B}^{n+1}={q_B}^n+\rho\frac{1}{2}({\bf F}^{-1}{\bf y})_{B},
\end{equation}
where 
\begin{equation}
y_{B}=({ {\bf T}}^t{\bf C}^{-1}{ {\bf C}}_{T,B}{\bf C}^{-1}
{ {\bf T}})
-{\rm Tr}({\bf C}^{-1}{ {\bf C}}_{T,B}),
\end{equation}
and 
\begin{equation}
F_{BB^{\prime}}=\frac{1}{2}{\rm Tr}({\bf C}^{-1}{{\bf C}}_{T,B}
{\bf C}^{-1}{ {\bf C}}_{T,B^{\prime}}).\label{fish}
\end{equation}
$C_{T,B}$ is the partial derivative of $C_T$ with respect to $q_B$. 
With eq.~[\ref{qb}], it is easy to see 
\begin{equation}
C_{T,B}=
\sum_\ell \frac{\partial C_T}{\partial D_\ell}\chi_{B\ell}\,.\label{ctb}
\end{equation}
The relaxation factor $\rho$ is a numerical constant between $0$ and $1$.
The converged power spectrum is independent of $\rho$, however 
setting $\rho < 1$ prevents the estimated $q_B$ from ``overshooting"
and wandering to some local extrema. Empirically, 
$\rho=0.5$ worked very well, and the $q_B$'s converged
after a few iterations.

\subsubsection{Decorrelated Band-Powers}\label{subsubsec:decorr}

Due to the nature of the experiment (partial sky coverage, correlated noise in 
the timestream, etc.), the band-power values are not independent. The 
correlations between them are given by the inverse of the Fisher matrix 
(eq.~[\ref{fish}]). To get the correct uncertainty estimate for a band-power value we 
either need to marginalize over all the other bands, or perform a linear 
transformation to decorrelate the data points. We chose
to decorrelate the data points to maintain as much information as 
possible. The decorrelation matrices are not unique:
our choice is the Hermitian square root of Fisher matrix, 
because it is localized and symmetric \citep{tegmark97b}. 
In the new basis, the local likelihood function is in canonical form. 

\subsubsection{Shape of the Likelihood Function}

It was pointed out by \citet{bond2000} that an offset log-normal 
function is a much better approximation to the likelihood function 
than a quadratic one. Failing to account for this 
could lead to serious misinterpretation of an experimental result, 
a most notable example being the quadrupole moment measured by COBE. 
In the decorrelated basis, the likelihood function near the maximum
($q_B={\overline q}_B$) can be approximated by
an offset log-normal function
\begin{equation}\label{entropy}
\ln{\cal L}({\bf q}) = \ln{\cal L}(\overline{{\bf q}})-\frac{1}{2}
\sum\limits_{B}\frac{(Z_B - \overline{Z}_B)^2}{\sigma_B^2}e^{2{\overline Z}_B},
\end{equation}
where the offset log-normal parameters ${\bf Z}$ are defined as
\begin{equation}\label{lognorm}
Z_B = \ln{(q_B + x_B)}.
\end{equation}
Even to the quadratic order, the Fisher matrix is only an approximation
to the curvature matrix of the likelihood function near the extrema.
To 
examine the real shape of the likelihood function, we 
explicitly fit the curvature $\sigma_B$ and log-normal offset $x_B$ 
and compared the results with the Fisher matrix 
(now diagonal) and the offsets suggested by \citet{bond2000} (see their 
eq. [28]). In general we found very good agreement.  In a few bands, however,
the uncertainty indicated by the true likelihood function 
is $\sim 15\%$ larger than that derived from the Fisher matrix. 
The uncertainty and log-normal offsets reported in this paper are determined
from fits to the likelihood function.

\subsubsection{Window Functions}\label{subsubsec:window}
Our goal is to produce a CMB power spectrum that can be used to 
constrain cosmological models.
Window functions are used to convert a given model power spectrum
to quantities directly comparable to the band-power measurements of the 
experiment. 
A window function $W_{B\ell}$ for band ``$B$'' should have the 
following property:
$$
\langle q_B \rangle=\sum_\ell (W_{B\ell}/\ell)D_\ell\,,
$$
where $q_B$ is the experimental band-power measurement. \citet{knox99}
derived the appropriate window function for the band-power quadratic 
estimator \citep{tegmark97b},
$$
W_{B\ell}/\ell=\frac{1}{2}\sum_{B'} (F^{-1})_{BB'}{\rm Tr}\left(
\frac{\partial C_T}{\partial D_\ell}C^{-1}C_{T,B'}C^{-1}\right)\,.
$$
This form of window function has been used in several other 
experiments \citep{halverson_thesis,pryke02,myers02}, and should serve as a good approximation 
for maximum likelihood band-power. 
Making use of eqs.~[\ref{fish}][\ref{ctb}], 
it can be shown that the pre-decorrelation window functions
satisfy the following normalization condition, 
\begin{equation}
\sum_\ell \chi_{B\ell}W_{B'\ell}/\ell=\delta_{BB'}\,.\label{norm}
\end{equation}
In Section~\S\ref{subsubsec:decorr}, 
we described the computation of a set of decorrelated 
band-powers from the Fisher matrix of the raw band-powers.
This same linear transformation is applied to the window functions
to produce a set of decorrelated window functions.

\begin{figure*}[t]
\centerline{
\psfig{figure=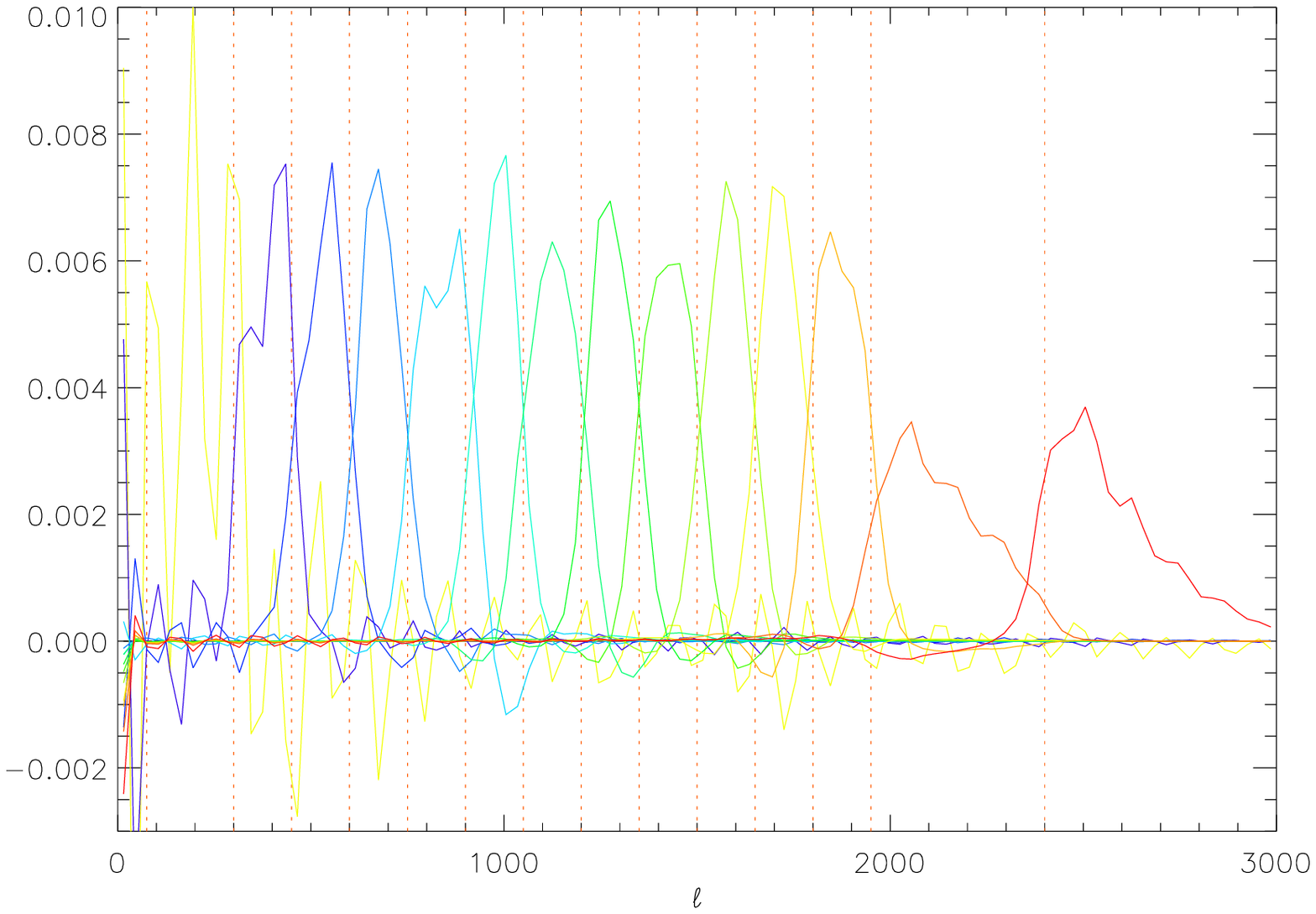,width=5in,height=3in,angle=0}}
\caption{The window functions ($W_{B\ell}/\ell$) for the 
decorrelated {\sc Acbar}
band powers.
The vertical lines show the band boundaries.
Numerical tabulations of these functions are given on the
{\sc Acbar} website.
}
\label{fig:window}
\end{figure*}

It is not necessary to calculate the window functions for all
$\ell$. However, we would like to sub-divide the $q_B$ bands into bins with 
finer $\ell$ space resolution than the $q_B$ themselves. Otherwise in order 
to satisfy the normalization condition eq.~[\ref{norm}], the pre-decorrelation
window 
function reduces to a Kronecker-$\delta$ in band index ``$B$'', or the tophat 
function given by $\chi_{B\ell}$. 
This is a reasonable approximation for experiments like BOOMERANG where 
the sky coverage, and therefore intrinsic $\ell$ resolution, is large enough to 
resolve all the structures in the power spectrum. 
For {\sc Acbar}, the $\ell$ resolution of the experiment is 
comparable to the expected structure in $D_\ell$ and it is essential
to precisely characterize the dependence of each band power on the 
details of the power spectrum.
The decorrelated window functions shown in Figure~\ref{fig:window} 
were calculated with 
a resolution in $\ell$ space of $\Delta \ell =30$.
The first window function has significant oscillations with 
$\Delta \ell \sim 110$.
This appears to be an result of the close spacing of the fields in the LMT
subtraction.
For sufficiently smooth theoretical models these oscillations will average
to a mean value, however, the exact dependence of the first band 
power on arbitrary input models could be complex. 
Numerical tabulations of the window functions are available on the {\sc Acbar} 
public web site.

\subsubsection{Monte Carlo Simulations}

\begin{figure*}[t]
\centerline{
\psfig{figure=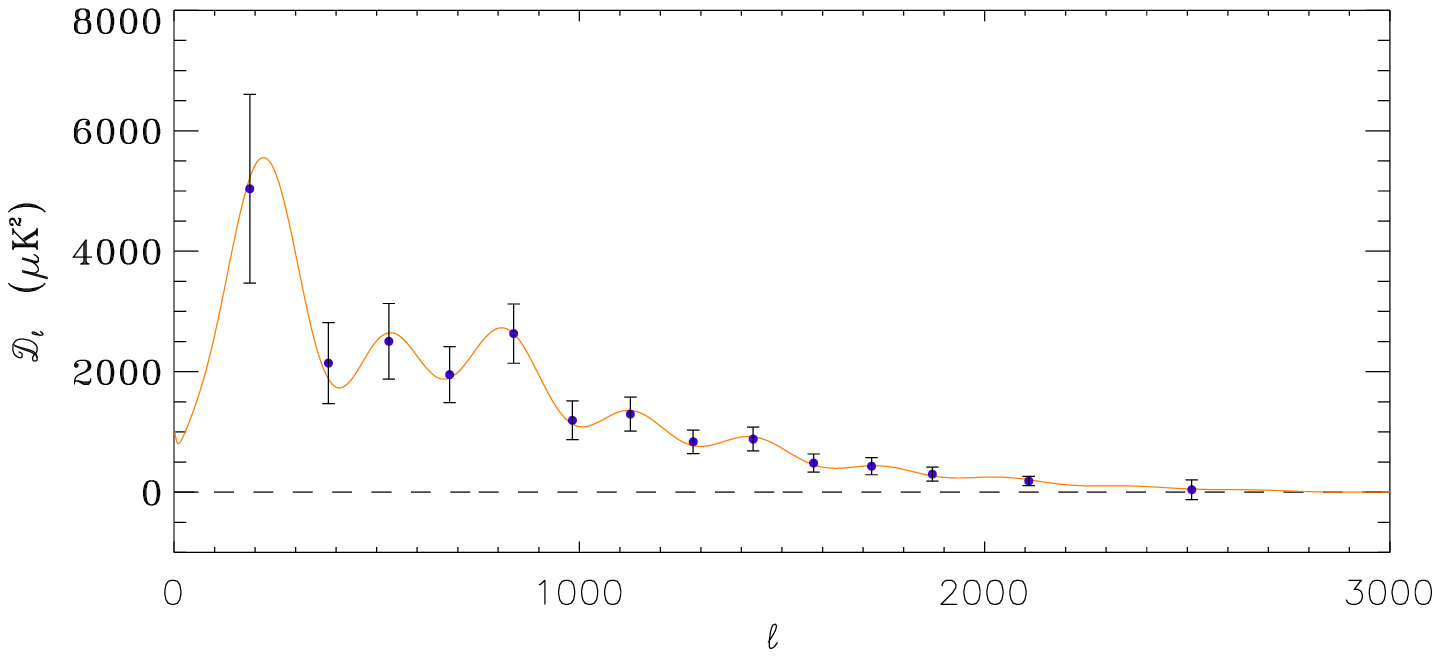,width=6in,height=3in,angle=0}}
\caption{The results of 300 Monte Carlo runs using the measured
{\sc Acbar} noise correlation. The solid line is the input fiducial
$\Lambda$CDM model. It is clear that the analysis method accurately
reproduces the input power spectrum.}
\label{fig:montecarlo}
\end{figure*}

In order to test our analysis technique, we performed 300 Monte Carlo
realizations of CMB2 and CMB5 maps that are consistent with our 
observed noise statistics
and the same underlying $\Lambda$CDM model.
These maps were processed exactly as the real data, including
using the same projection of corrupted modes, in order to
test the ability of our pipeline to accurately determine the
input CMB power spectrum. A quadratic estimator technique 
\citep{tegmark97b} is used to rapidly determine the
power spectrum for each of 300 realizations.
In Figure~\ref{fig:montecarlo}, we show the resulting average power spectrum
overlaid on the input model.
It is clear that the input model is recovered without bias, even in
the low $\ell$ bins that are sensitive to the angular scales on which
the mode removal is occurring.

\section{Results}\label{sec:results}

\subsection{Maps}
\begin{figure*}[t]
\centerline{
\psfig{figure=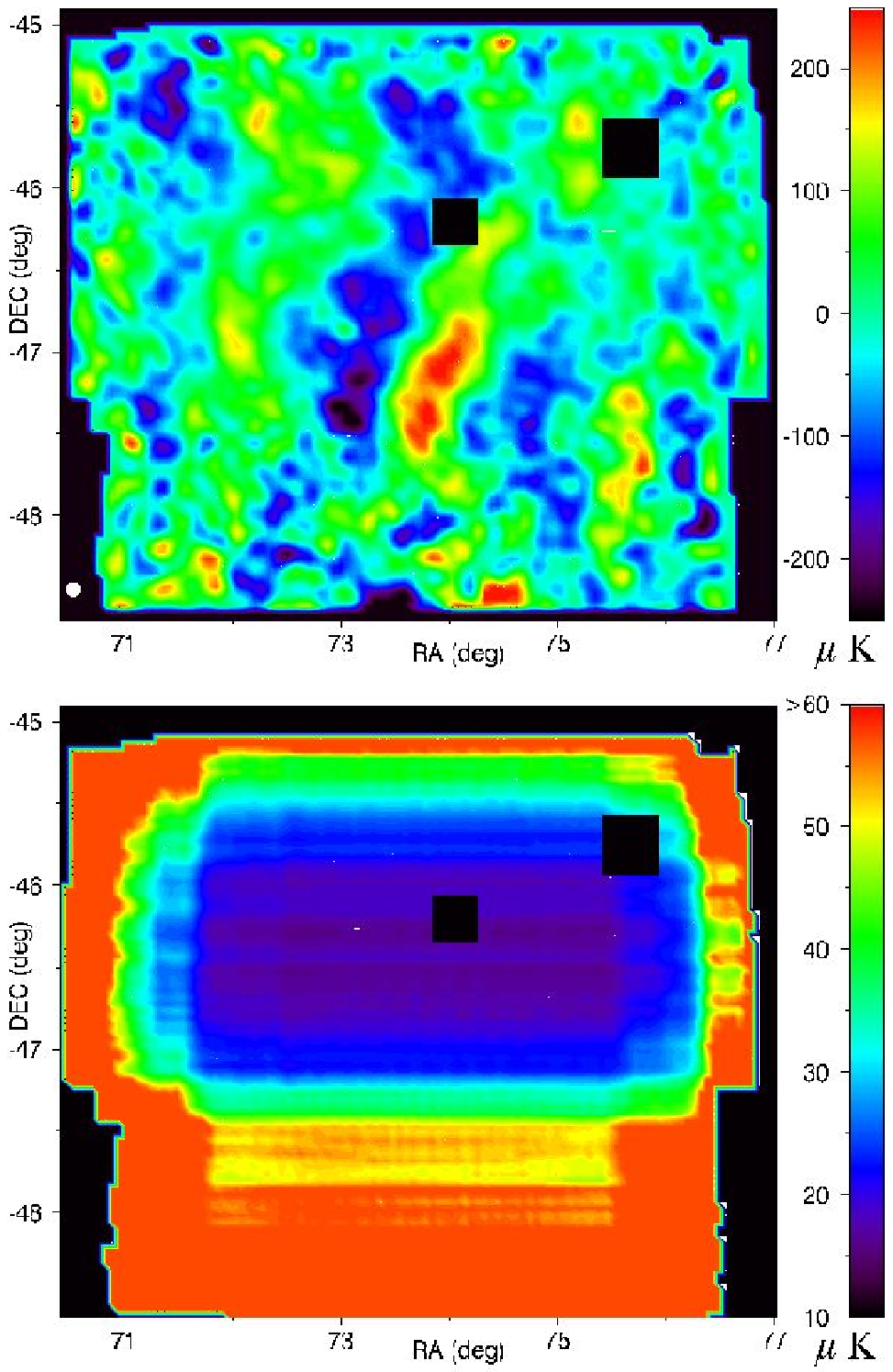,width=4in,height=6in,angle=0}
}
\caption{The top panel shows the LMT differenced, atmospheric mode removed,
noise weighted, coadded map for the CMB2 field.
The guide quasar and radio source Pictor-A have been replace with black pixels.
The small white circle in the lower left hand corner of the map
represents the FWHM of the average array element beamsize as determined from
the coadded quasar image.
The predominance of extended structure in the vertical direction is because
the extended horizontal structure has been projected out in the atmospheric
mode removal.
The lower panel shows the noise in the LMT difference map as a function of
position.}
\label{fig:cmb2_map}
\end{figure*}

\begin{figure*}[t]
\centerline{
\psfig{figure=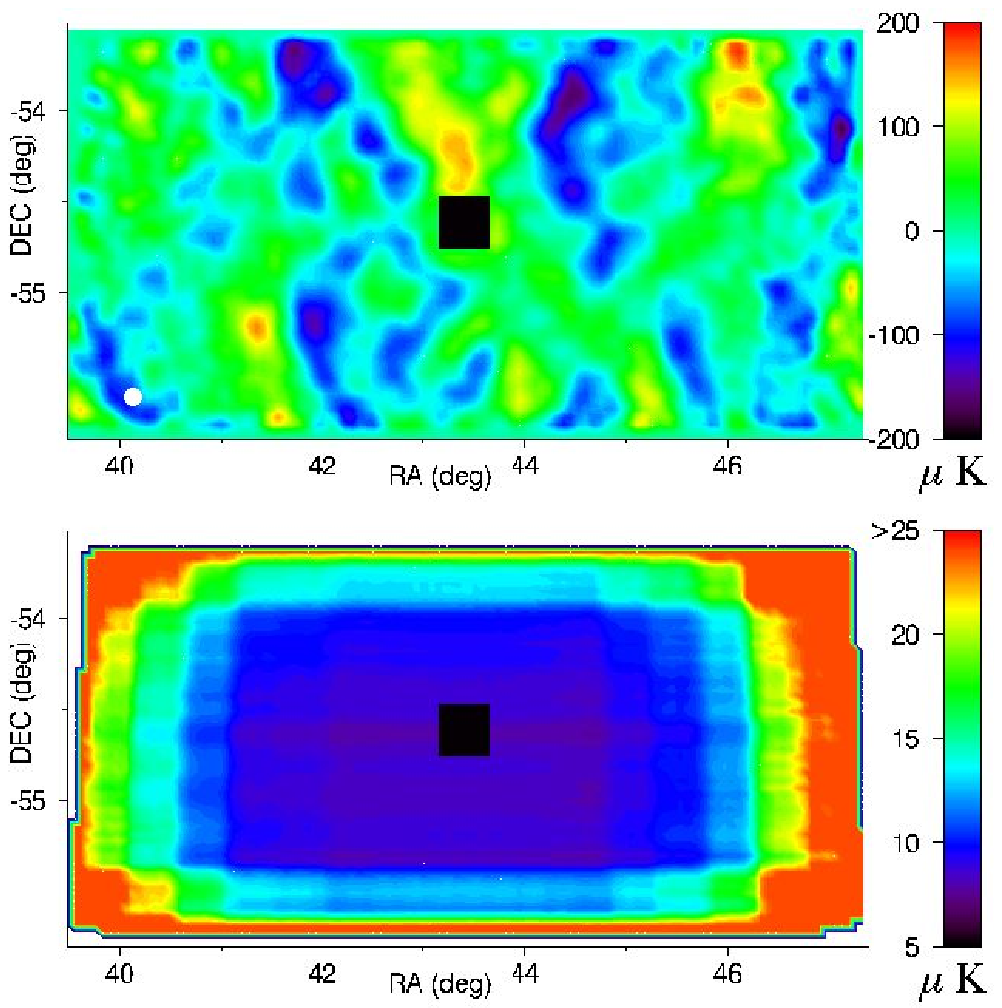,width=4in,height=4in,angle=0}
}
\caption{The same maps as shown in Figure~\ref{fig:cmb2_map}, but
for the CMB5 field.
The signal-to-noise in the center of this map for degree-scale
fluctuations approaches 100.
}
\label{fig:cmb5_map}
\end{figure*}

Using the technique described in \S~\ref{subsubsec:nwcm}, the LMT subtracted 
data are combined to form the noise weighted coadded maps shown in 
Figures~\ref{fig:cmb2_map} and \ref{fig:cmb5_map}. 
Due to computational limitations, we choose the map pixelization to be
$2.5^{\prime}$.  
This is a compromise that allows us to recover the power spectrum
with minimal loss of information while keeping the time and memory requirements
reasonable for a single 16 processor, 64 GB RAM node of the National Energy 
Scientific Computing (NERSC) IBM SP computer. 
These maps have had the two guide quasars and one additional detected radio
source (Pictor A) removed; black pixels mark the effected areas. 
Pictor A is observed to be extended and we remove a larger 
number of pixels around the position of this source.
These maps have not yet had the undetected PMN source catalog and IRAS dust
templates projected out; however, no other significant sources remain. 
For the purposes of presentation, we have smoothed the pixelized map
with a $4.5^{\prime}$ FWHM Gaussian beam. 
The lower panels in Figures~\ref{fig:cmb2_map} and \ref{fig:cmb5_map} show the 
RMS noise in the LMT difference maps as a function of position.
Due to differences in map coverage, the noise varies across the map;
in the central region, the RMS noise per $5'$ beam is found to be
$17\,\mu$K and $8\,\mu$K for CMB2 and CMB5 respectively. 
On degree angular scales, the S/N in the center of the CMB5
map approaches 100.  

\subsection{Band-Powers}
After projecting out the PMN catalog and dust templates, the formalism
described in \S~\ref{subsec:power} is used to to determine the
angular power spectrum. 
The resulting band-powers are correlated at the $\sim 20\%$ level before they
are decorrelated.
The decorrelated band power values are plotted in Figure~\ref{fig:bandpower} with 
a fiducial $\Lambda$CDM model; these same results are listed in 
Table~\ref{tab:bands}.
The corresponding decorrelated band-power window functions are plotted in 
Figure~\ref{fig:window}.
Before decorrelation, the window functions are typically 
anticorrelated to those of adjacent 
bands at a level of $\sim 15\%$. After decorrelation, the anticorrelations
decrease substantially.
The decorrelated band-powers and tabulated window functions are available on the 
{\sc Acbar} experiment public 
website\footnote{\tt http://cosmology.berkeley.edu/group/swlh/acbar}.

\begin{table*}
\caption{\label{tab:bands}Joint Likelihood Bandpowers}
\small
\begin{center}
\begin{tabular}{ccccccc}
\hline\hline
\rule[-2mm]{0mm}{6mm}
$\ell$ range & $\ell_{-}$ & $\ell_{+}$& $\ell_{\rm eff}$
& $q$ ($\mu{\rm K}^2$)& $\sigma$ ($\mu{\rm K}^2$) &
x ($\mu{\rm K}^2$)\\
\hline
75-300 & & & 187 & 6767 & 1323 & -45 \\
300-450 & 307 & 459 & 389 & 2874 & 605& -501\\
450-600 & 462 & 602 & 536 & 2716 & 498& -328\\
600-750 & 615 & 744 & 678 & 2222 & 360 & -391\\
750-900 & 751 & 928 & 842 & 2300 & 355 & -154\\
900-750 & 921 & 1048 & 986 & 798 & 153 & -76 \\
1050-1200 & 1040 & 1214 & 1128 & 1305 & 208 & -10\\
1200-1350 & 1207 & 1352 & 1279 & 583 & 130 & 36\\
1350-1500 & 1338 & 1513 & 1426 & 628 & 134 & 111\\
1500-1650 & 1510 & 1649 & 1580 & 351 & 110 & 160\\
1650-1800 & 1648 & 1785 & 1716 & 248 & 99 & 149\\
1800-1950 & 1782 & 1953 & 1866 & 361 & 132 & 277\\
1950-2400 & 1946 & 2227 & 2081 & 337 & 109 & 557\\
2400-3000 & 2377 & 2651 & 2507 & 307 & 275 & 1888\\
\hline
\end{tabular}
\end{center}
\tablecomments{Decorrelated band-powers $q_B$, uncertainty 
$\sigma_B$
, and log-normal offset $x_B$ from the 
joint likelihood analysis of CMB2 and CMB5.
The PMN radio point source and IRAS dust foreground templates have been
projected out in this analysis.
This is the same data shown in Figure~\ref{fig:bandpower}.
$\ell_\pm$ are the multipole $\ell$ where the window functions drop to
half of their peak values; the effective multipole of the band is 
$\ell_{\rm eff}\equiv \sum \ell (W_\ell/\ell)/ \sum (W_\ell/\ell)$,
where the average is done for $\ell_{-}<\ell<\ell_{+}$. The window
function for the first band is oscillatory, and $\ell_{\rm eff}$ is
taken to be the average of the upper and lower $\ell$ boundary.}

\normalsize

\end{table*}

\begin{table*}
\caption{\label{tab:foreground1}CMB2 Foreground modes}
\small
\begin{center}
\begin{tabular}{cccc}
\hline\hline
\rule[-2mm]{0mm}{6mm}
& RA\tablenotemark{a} & DEC\tablenotemark{a} & Degrees of Freedom\\
PMN J0455-4616   & $04^h55^m51^s$ & $-46^\circ 15'59^{\prime\prime}$&  49 \\
Pictor A    & $05^h19^m50^s$ & $-45^\circ 46'44^{\prime\prime}$&  81\\
SFD dust map\tablenotemark{b} & - & - & 1\\
Other PMN sources &-&-& 102\\
\hline
\end{tabular}
\tablenotetext{a}{Epoch 2000.}
\tablenotetext{b}{Predicted $150$GHz emission from DIRBE - calibrated IRAS 100
$\mu m$ map \citep{finkbeiner99}.}
\tablecomments{These modes are removed from the CMB2 field and do not
contribute to the power spectrum.}
\end{center}
\normalsize
\end{table*}

\begin{table*}
\caption{\label{tab:foreground2}CMB5 Foreground modes}
\small
\begin{center}
\begin{tabular}{cccc}
\hline\hline
\rule[-2mm]{0mm}{6mm}
& RA & DEC & Degrees of Freedom\\
PMN J0253-5441   & $02^h53^m29^s$ & $-54^\circ 41'51^{\prime\prime}$&   49\\
 SFD dust map& - & - & 1\\
Other PMN sources &-&-&68 \\
\hline
\end{tabular}
\tablecomments{Same as table~\ref{tab:foreground1}, for CMB5 field.}

\end{center}
\normalsize
\end{table*}

\section{Foregrounds}\label{sec:foregrounds}

Sources of foreground emission have the potential to contaminate 
the measured CMB power spectrum. 
Multifrequency observations can be used
create templates for some foregrounds and remove their effect 
from the power spectrum.
At our observation frequencies, thermal emission from galactic dust,
extragalactic radio sources, dusty protogalaxies, and the Sunyaev-Zel'dovich 
effect are all potential sources of foreground emission \citep{tegmark99}. 

The observed CMB fields were chosen to lie in a region of minimal dust contrast
near the southern galactic pole.
Templates for the dust emission are determined by performing a LMT 
difference on the IRAS/DIRBE maps published by \cite{schlegel98}.
The emission is extrapolated to $150\,$GHz, using the scaling of 
\cite{finkbeiner99}. 
From this analysis, we expect contributions to the observed 
temperature anisotropy power in the differenced maps of 
$9\,\mu{\rm K}^2$ and $70\,\mu{\rm K}^2$ for CMB2 and CMB5 respectively.
This power is dominated by structure on the scale of the map and 
decreases on smaller angular scales ($\propto 1/\ell$).
The contribution of dust to the observed power is therefore expected to be 
completely negligible across the entire {\sc Acbar} angular power spectrum. 
Projecting out dust emission using the IRAS/DIRBE template 
results in the loss of only one degree of freedom and
has no significant effect on the resulting power spectrum.

\begin{figure*}[t]
\centerline{
\psfig{figure=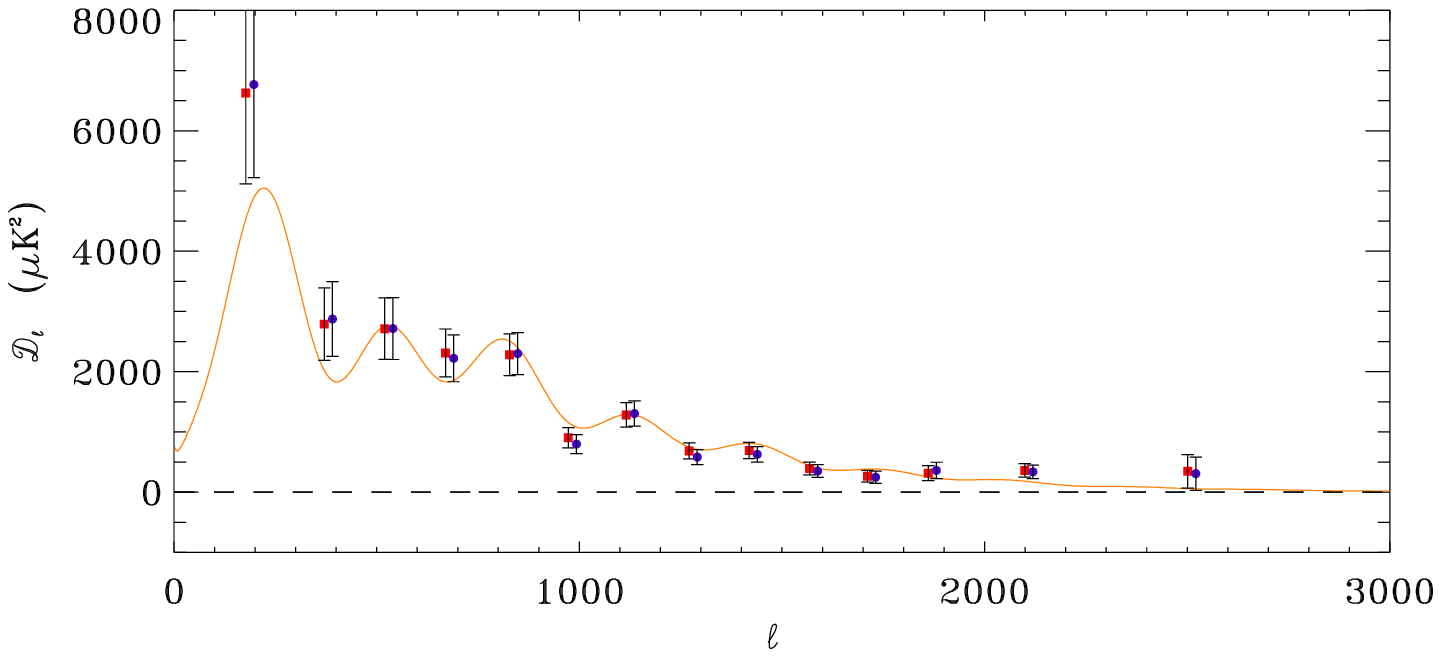,width=6in,height=3in,angle=0}
}
\caption{The CMB angular power spectrum with (circle) and without 
(square) the
projection of templates for PMN radio point sources and thermal dust.
Removing these sources of foreground emission has no significant
effect on the measured power spectrum.
The bandpower values and confidence intervals for the foreground projected
power spectrum are listed in Table~\ref{tab:bands}.
}
\label{fig:bandpower}
\end{figure*}

Extragalactic Radio sources are another potential source of foreground confusion.
By comparing the positions of known radio sources from the PMN $5.0\,$ GHz 
survey \citep{wright94} with the CMB maps, we have detected 3 radio sources with 
significance greater than $3 \sigma$.
Two of these are the guiding quasars deliberately selected to lie in the 
center of the observed fields.
In Tables~\ref{tab:foreground1} and \ref{tab:foreground2}, 
we list the coordinates
of these sources and the number of pixels removed to clean them from the maps.
It is possible that many PMN sources lie just below the threshold 
of detection and therefore contribute statistically to the observed power spectrum.
We created templates which include all 170 PMN point sources in the CMB 
field with fluxes greater that $40\,$mJy at $5.0\,$GHz and use it to project 
out any possible contribution to the measured power spectrum.
In Figure~\ref{fig:bandpower}, we show the CMB angular power spectrum with and 
without the projection of the undetected point sources and the dust templates.
Unless stated otherwise, all results in this paper are found with an
analysis that removes these source templates.

\begin{figure*}[t]
\centerline{
\psfig{figure=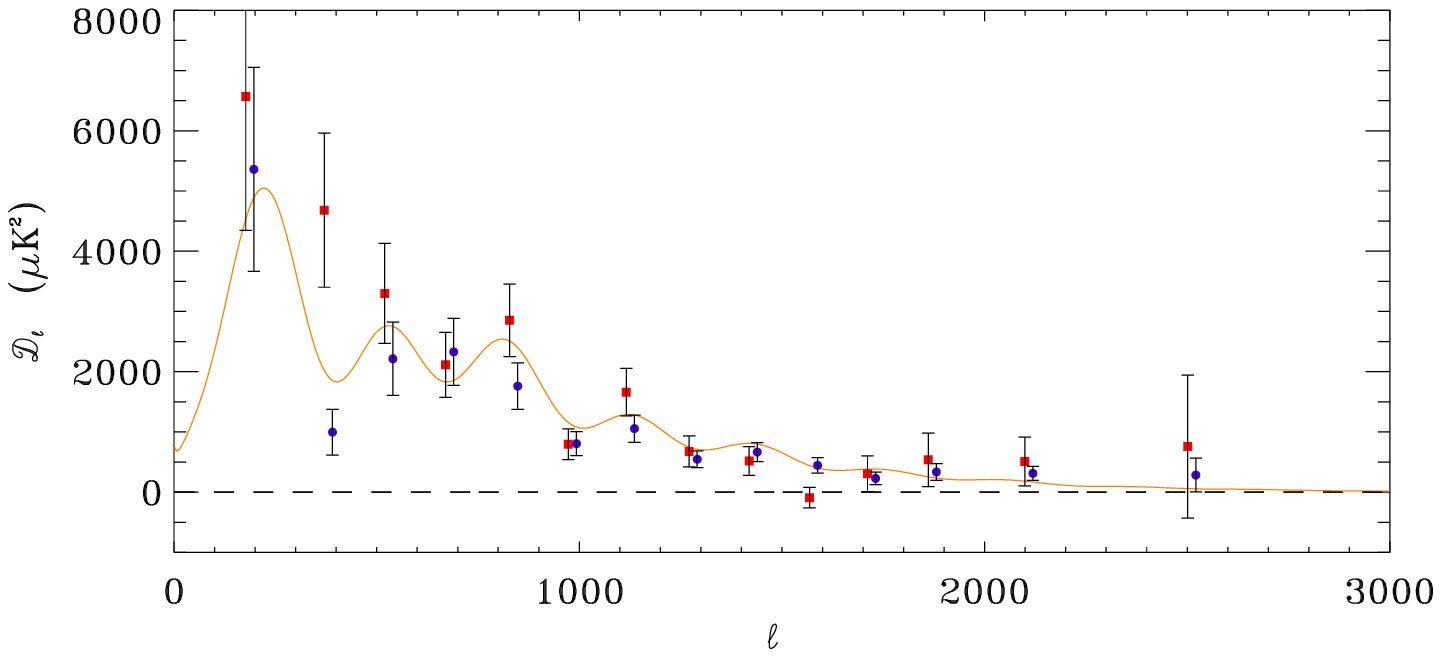,width=6in,height=3in,angle=0}
}
\caption{The CMB angular power spectrum derived from CMB2 (square) and
CMB5 (circle) field.
}
\label{fig:foreground}
\end{figure*}

It is also possible that dusty protogalaxies could contribute to the 
observed CMB signal.
Unlike the radio sources, we have no template for the positions of 
these objects. 
\cite{blain98} have estimated the contribution of dusty protogalaxies to 
measurements of CMB anisotropy and predict a contribution to the $150\,$GHz
power spectra of $(\Delta T)^2 \sim 10\, (\ell/2500)^2\, \mu{\rm K}^2$.
Therefore, dusty protogalaxies are not expected to contribute
significant signal to the observed CMB anisotropy.

The most likely candidate for contamination of the observed {\sc Acbar}
power spectrum is the Sunyaev-Zel'dovich (SZE) effect in distant clusters 
of galaxies.
Analytic models and numerical simulations generally predict a SZE power spectrum with 
a broad peak at angular scales corresponding to $\ell \sim 10^4$ \citep{white02}.
Deep observations with the Cosmic Background Imager (CBI), observing 
in a frequency band of $26-36\,$GHz and at angular scales corresponding to 
$\ell=2000-3500$, have reported a significant detection of 
excess power $(\Delta T)^2=506^{+116}_{-149}\,\mu {\rm K}^2$ \citep{mason02}.
Interpreting this power as the SZE leads to a value for the RMS mass 
fluctuations inside spheres of $8\, h^{-1}$Mpc of $\sigma_8 \sim 1.0$, close 
to the upper limits allowed by observations of the matter power 
spectra \citep{bond02,komatsu02}.
Similar observations with the Berkeley Illinois Maryland Association (BIMA) array 
at a frequency of $28.5\,$GHz and on angular
scales closer to the expected peak in the SZE power spectrum 
$(\ell_{\rm eff}= 6864)$ also find excess power, 
$(\Delta T)^2=202^{+139}_{-135}\,\mu{\rm K}^2$ {\citep{dawson02}}
that is marginally consistent with the results reported by CBI.
If these signals are in fact due to the SZE, we can use the spectral 
dependence of the SZE to predict the contribution to the {\sc Acbar}
power spectrum. 
Neglecting relativistic corrections, the observed change in CMB temperature
in the direction of a galaxy cluster with Comptonization $y$ is given by 
\begin{equation}
\Delta T= T_{CMB}\, y \left(\frac{x(e^x+1)}{e^x-1}-4\right)\,, 
\end{equation}
where $x=h\nu/kT_{CMB}$.
Therefore, the CMB temperature anisotropy power due to the SZE should be a factor
$\Delta T^2 (150\, {\rm GHz})/\Delta T^2 (30\, {\rm GHz})=0.238$ 
smaller for {\sc Acbar} than was found by CBI and BIMA.
In Figure~\ref{fig:bandpower} we show the {\sc Acbar} and CBI power spectra 
plotted on top of a $\Lambda$CDM power spectrum. 
The marginal detection of excess power in the last {\sc Acbar} band-power 
is consistent with the extrapolation of the CBI excess power to $150\,$GHz 
if the signal is due to SZE.
However, due to the large uncertainty on this point, the {\sc Acbar}
results are also consistent with the origin of the CBI excess being 
due to CMB anisotropy from local structure and non-standard inflationary 
models \citet{cooray02,griffiths02} or a new population of faint radio point 
sources.

\section{Systematic Tests}\label{sec:systematic}

\subsection{Gaussianity}\label{subsec:gaussianity}

The derivation of the band-power ${\bf q}$ is based on the assumption that the 
signal is Gaussian distributed. 
This assumption is tested by examining the distribution of signal-to-noise 
eigenmodes ${\bf T}_s$ (Appendix \ref{app:sn}). After the maximum likelihood 
band-powers are found,
the total covariance matrix (theory and noise) in the signal-to-noise basis 
is diagonalized 
and normalized, such that in the new basis the covariance matrix becomes 
an identity matrix. Then
the data vector ${\bf T}_s$ is transformed accordingly and compared to 
a Gaussian function with unit variance.
When the 2133 high signal-to-noise modes are compared to the Gaussian 
distribution, the Kolmogorov-Smirnov statistic 
is determined to be $D=0.0159$.
The corresponding probability for values drawn from a Gaussian distribution 
to exceed this value is $P_{ks}(D>0.0159)=64.7\%$, and 
the distribution of eigenmodes is determined to be consistent with Gaussian.
This result justifies the use of the band-power estimation algorithm, 
however, we point out that this is not a sensitive test for any original 
non-Gaussianity in the map, such as
skewness introduced by the SZE or other secondary anisotropies. 
The LMT differencing scheme reduces the 
skewness, and then the arbitrary eigenvector directions in the 
signal-to-noise eigenmode transformation further dilutes 
any residuals. 
To test the inflationary prediction that the primary 
CMB anisotropy is Gaussian distributed and to search for non-Gaussian 
signatures due to secondary anisotropies, a reanalysis of the data is required.

\subsection{Difference Map Power Spectrum}\label{subsec:jackknife}

We performed a number of tests for systematic errors in the {\sc Acbar} CMB power 
spectrum.
These tests include ``jackknife'' tests where the raw data for each field 
is divided 
into two sets that can be subtracted to form a difference map.
The power spectrum derived from the difference map is then examined
for significant departures from zero that would signal the presence of
systematic differences between the two halves of the data set. 
This technique provides a powerful method to determine if ground pickup, 
calibration, detector time constants, or other potential 
systematic effects contribute significantly to the observed CMB power 
spectrum.  The difference map band-powers at high $\ell$ are also sensitive to 
errors in the noise estimate. A positive/negative residual 
in the highest $\ell$ band would indicate an underestimate/overestimate of  
experimental noise.

Because of the adaptive mode removal scheme described in \S~\ref{subsubsec:nwcm}, 
the differenced power spectrum is not expected to be strictly zero. 
If ${\bf K}_1$ and 
${\bf K}_2$ are the correction matrices defined in eq.~[\ref{k}] for two 
maps ${\bf T}_1$ and ${\bf T}_2$, the differenced map 
$({\bf T}_1-{\bf T}_2)/2$ will have an expectation value of 
$({\bf K}_1-{\bf K}_2){\bf T}^0/2$. 
How completely the maps subtract depends on the matrix $({\bf K}_1-{\bf K}_2)$.  
Monte Carlo simulations are used to produce 300 unique realizations of the sky
(assuming a $\Lambda$CDM Universe) and noise consistent 
with that measured by the experiment. Difference maps are generated 
for these realizations using the ${\bf K}$ matrices determined from the
data. The residual difference power spectrum and uncertainty are calculated 
for the Monte Carlo simulations and compared to the observed residuals.
Typically, the residual due to mode removal is only 
significant when differencing data taken under very different atmospheric
conditions, and then affects only the lowest $\ell$ bin. 

\begin{figure*}[t]
\centerline{\psfig{figure=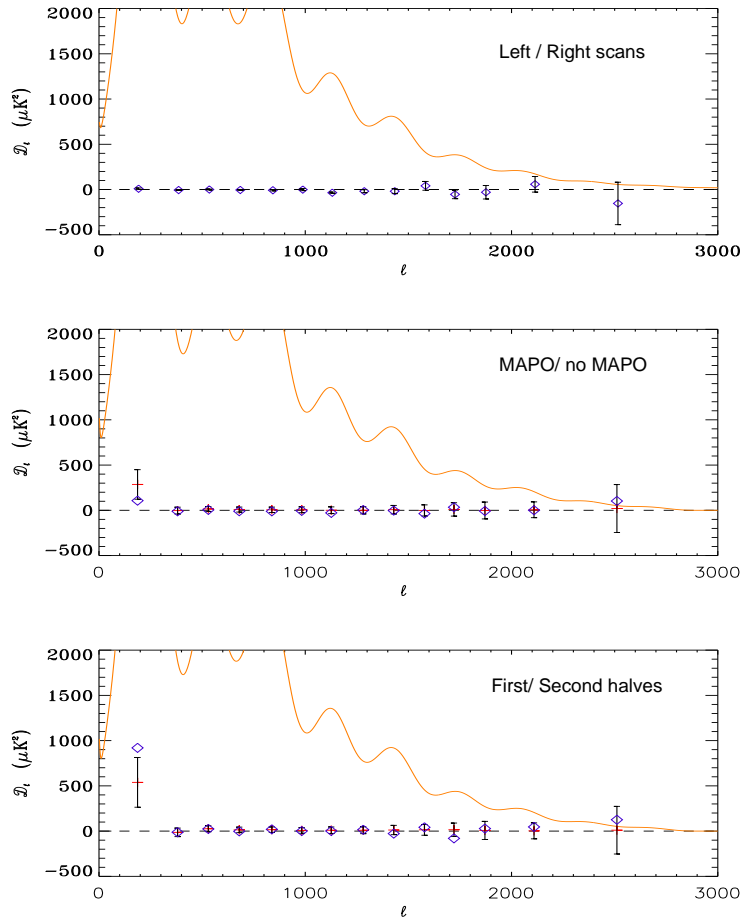,width=4in,height=5in,angle=0}}
\caption{Power spectra of difference maps produced from three
different data jackknifes.
The top panel shows the power spectrum produced from the difference
of maps made with left and right going chopper sweeps.
The central panel shows the power spectrum produced from the difference
of maps made with values of azimuth toward and away from the MAPO building.
The bottom panel shows the power spectrum produced from the difference
of maps made from the first and second halves of the data for each of the
observed fields CMB2a, CMB2b, CMB5a, and CMB5b.
In the central and bottom panels the predicted residuals (cross) and 
uncertainties
are determined by 300 Monte 
Carlo simulations of
each difference map and include effects of incomplete subtraction
due to atmospheric mode removal.
In all cases, the difference band-powers (diamond) are consistent with there 
being no significant
signal in the subtracted maps. These tests indicate that the noise estimate
is accurate and the band-power results are free from significant systematic
errors. 
}
\label{fig:jackknife}
\end{figure*}

Two maps were created by separating the data into two halves 
corresponding to the left-going and right-going portions of each chopper sweep. 
The power spectra of the difference between these maps should be sensitive 
to systematic errors due to chopper induced microphonic response or detector 
time constants. 
Because the left and right going sweeps are carried out in exactly
the same atmospheric conditions and are subject to the same mode
removal, there is no expected residual in the subtracted maps.  
The power spectrum derived from the L/R differenced data is shown in 
the top panel of Figure~\ref{fig:jackknife}. 
The error bars on the differenced map power spectra are derived from the 
Fisher matrix. 
The resulting power spectrum is consistent with there being no
systematic difference between the maps. 

Next, the data were separated by the azimuth of the observations
into two halves corresponding to observations with the telescope pointing 
in the $180^\circ$ azimuth ranges toward and away from the nearby Martin A. 
Pomerantz Observatory (MAPO) building.
The MAPO building is the largest object on the otherwise smooth horizon
seen by {\sc Acbar} at the Pole and is expected to be the largest source of 
systematic signal due to the modulation of the far sidelobes of the 
{\sc Acbar} beams. 
Therefore, the power spectrum for the differenced map is 
a sensitive test for ground pickup that is not removed by the LMT 
differencing.
The resulting power spectra is plotted in the middle panel 
of Figure~\ref{fig:jackknife}; the results are consistent with the 
the residual calculated from the Monte Carlo simulation. 

In order to test for systematic changes in pointing, calibration or beamsize
over the course of the observations, we also create a difference map
by separating the data in time.
As described in \S~\ref{sec:observations}, each of the CMB fields observed 
actually consists of two subfields 
offset from one another by $0.5^{\circ}\sec\delta$.
Due to changes in beamshape across the chopper swing, there are subtle
differences between the subfield maps that could lead to a non-zero residual 
signal when these maps are subtracted.
However, when the maps are added, the resulting beams are represented
by the sum of the beams in the input maps and the power spectrum
should be unbiased.
In order to avoid these complications, differenced maps 
are created from the first and second halves of the data
for each of the four observed subfields CMB2a, CMB2b, CMB5a, and CMB5b.
Because the data from which the maps are produced are well separated 
in time and are subject to different atmospheric mode removal, 
the predicted residuals for the subtracted maps can be significant
on large angular scales. 
The lower panel of Figure~\ref{fig:jackknife} shows the deference map power 
spectrum with the Monte Carlo determined residuals and uncertainties;
again, there is no evidence for significant excess power in the difference map.

\begin{figure*}[t]
\centerline{
\psfig{figure=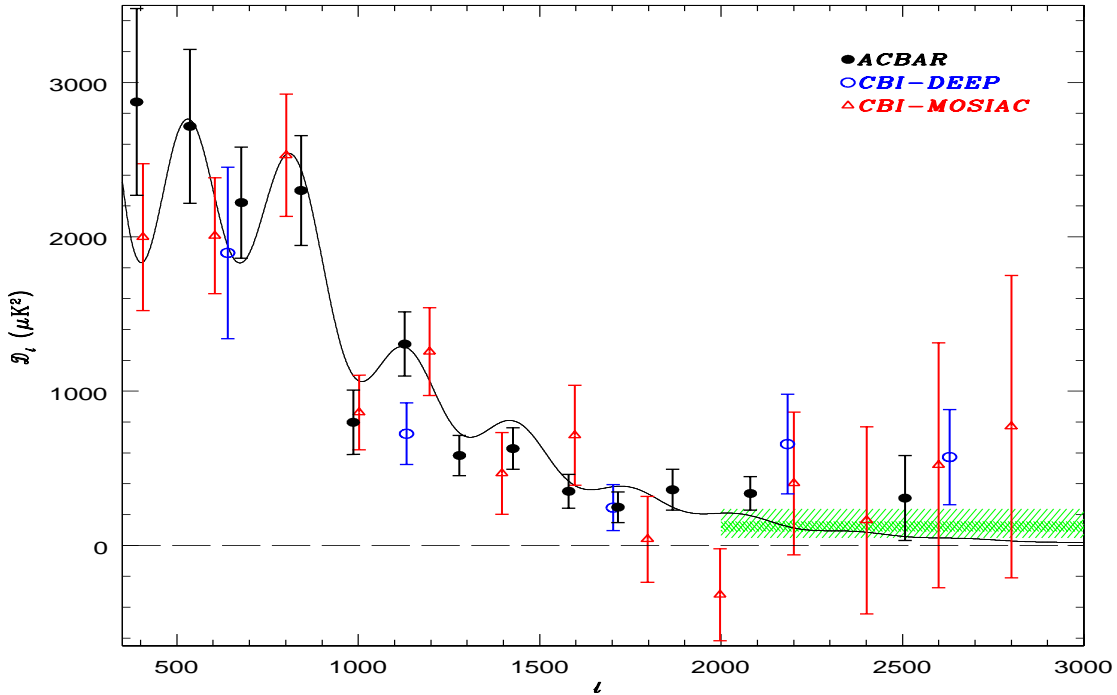,width=6in,height=4in,angle=0}
}
\caption{The CMB angular power spectrum 
for the CBI-Deep, CBI-Mosaic, and {\sc Acbar} experiments.
The data are plotted on top of a fiducial $\Lambda$CDM model.
The shaded green bar shows the expected contribution to the ACBAR power 
spectrum if the excess power found in the CBI-Deep data is due to
the SZE. 
The ACBAR data are consistent with the CBI excess being due to SZE, but 
provide no signifincant new constraints on the source of the signal.
}
\label{fig:compare}
\end{figure*}

\section{Conclusion}\label{sec:conclusion}

We have used the {\sc Acbar} receiver to measure
the angular power spectrum of the CMB at a frequency of $150\,$GHz
over the multipole range $\ell = 100-3000$. 
The power spectrum we present is derived from approximately 21 weeks 
of Austral winter observations with {\sc Acbar} installed on the $2.1\,$m
Viper telescope at the South Pole.

In the course of analyzing this data, we have employed new analysis 
techniques designed
specifically for high sensitivity ground based CMB observations.
Monte Carlo simulations are used to verify that the analysis method 
accurately recovers the power spectrum without bias.
The power spectrum we present is robust and has passed stringent tests
for systematic errors.
Galactic dust emission and radio point sources do not contribute 
significantly to the observed power and are projected
out in the final analysis.
Although dusty protogalaxies cannot be ruled out as a source of confusion,
the expected contribution to the measured power is negligible.
Overall, the resulting power spectrum appears to be consistent with the
damped acoustic oscillations expected in standard cosmological models.
In a companion paper \citep{goldstein02}, the {\sc Acbar} power spectrum
is used to place constraints on cosmological models.
The power in the highest $\ell$ {\sc Acbar} bin is consistent with the
excess power measured in the Deep CBI pointings. 
At this point, the {\sc Acbar} data lack the 
sensitivity to place significant constraints on the origin of the 
excess power observed by CBI. 

We acknowledge assistance in the design and construction of {\sc Acbar} by the 
UC Berkeley machine and electronics shop staff. 
The support of Center for Astrophysics Research in Antarctica (CARA) 
polar operations has been essential in the installation and operation of 
the telescope.
Percy Gomez, Kathy Romer, and Kim Coble are thanked for their assistance 
in monitoring the observations and telescope pointing.
We thank Nils Halverson, Julian Borrill, and Radek Stompor for a careful 
reading of the draft and useful comments on analysis algorithms.
Finally, we would like to thank John Carlstrom, the director of CARA, 
for his early and continued support of the project.
The ACBAR program has been primarily supported by NSF office of polar
programs grants OPP-8920223 and OPP-0091840.
This research used resources of the National Energy Research Scientific 
Computing Center, which is supported by 
the Office of Science of the U.S. Department of Energy under Contract 
No. DE-AC03-76SF00098. 
Chao-Lin Kuo acknowledges support from a Dr. and Mrs. CY Soong fellowship
and Marcus Runyan acknowledges support from a NASA Graduate Student 
Researchers Program fellowship.
Chris Cantalupo, Matthew Newcomb 
and Jeff Peterson acknowledge partial financial support 
from NASA LTSA grant NAG5-7926.

\appendix

\section{A. Noise Estimate}\label{app:noise}

Each {\sc Acbar} data file contains observations at $n_d$ pointings in 
declination.
Observations at each declination consist of lead, main, and trail
pointings in RA which we refer to as {\it stares}.
Each {\it lead} or {\it trail} stare has $2n_s$ 
($n_s$ left going and $n_s$ right going) chopper sweeps, 
while the {\it main} stares have $4n_s$ sweeps. 
Each sweep contains $n_x$ samples of the detector signals. 
The goal is to estimate the noise correlation in the 
$M-(L+T)/2$ differenced, coadded maps for each file, and then use this 
noise correlation in eq.~[\ref{cn}] to derive the final pixel space 
correlation matrix ${\bf C}_N$. 
In order to get better statistics for the noise determination,
the noise for each sweep is calculated before the sweeps are coadded so
that there are as many realizations as possible.

A file is divided into $n_d\times n_s$ sets of sweeps, each containing $(2,4,2)$ 
sweeps in the (L,M,T) fields on $n_d$ strips in declination. 
The sweeps in the sets of stares are grouped by their time order so that 
the first sets contains the first sweeps from each stare and the last set 
contains the last sweeps.
The sets of sweeps are combined (using $M-(L+T)/2$) to form a $n_x$-component 
data vector for each set of sweeps. 
The correlations are computed from the LMT subtracted sets of sweeps.

The data for one channel is now reduced to the array 
{\tt data[}$n_x$,$n_d$,$n_s${\tt ]} which we write as $D_{i,j,k}$. 
Assuming that the data are dominated by noise, the following moments can be 
determined
\begin{align*}
m_1(i,j)&=\langle D_{i,\ell, k}\times D_{j,\ell, k} \rangle_{\{\ell,k\}},\\
m_2(i,j,\Delta s)&=\langle D_{i,\ell, k}\times D_{j,\ell, k+\Delta s}
\rangle_{\{\ell,k\}},\\
m_3(i,j,\Delta d,\Delta s)&=
\langle D_{i,\ell, k}\times D_{j,\ell+\Delta d, k+\Delta s} 
\rangle_{\{\ell,k\}}.
\end{align*}
They are interpreted as correlations within a data strip ($m_1$), among
data strips in a constant declination ($m_2$), and among different 
declinations ($m_3$). These are all correlations within the same channel. 
There are also cross-channel correlations for $m_1,$ etc., 
that are also largely due to the atmosphere.

As was pointed out in \S\ref{sec:analysis}, the pixel space correlation matrix 
is simply the weighted average of ${\bf \Pi}{\bf N}{\bf \Pi}^t$, or 
$\langle \tilde {\bf d}\tilde {\bf d}^t\rangle$. 
One could calculate it directly from ``cleaned'' timestream $\tilde {\bf d}$, 
however after the mode removal, the correlation within a sweep is not just a 
function of $|i-j|$ anymore, instead it is a function of both $i$ and $j$. 
This prevents the application of Wiener-Khinchin theorem (PSD 
$\leftarrow{FFT}\rightarrow$ ACF) and would make the algorithm
time consuming. 
Instead, the correlation is determined before the projection destroys 
the ``stationary" property of the noise by calculating 
${\bf N}=\langle {\bf dd}^t\rangle$ first, then using 
$\langle \tilde {\bf d}\tilde {\bf d}^t\rangle=
{\bf \Pi}\langle {\bf dd}^t\rangle{\bf \Pi}^t$.
In Appendix~\ref{app:fft}, the application of the FFT for piecewise 
stationary random processes such as the {\sc Acbar} noise is described. 
A PSD is by definition positive. Due to limited numbers of realizations
the PSD calculated from the ACF is not always positive. We therefore
parametrize the PSD with the first three terms of the Laurent series, 
$$
PSD=a_0+a_1f^{-1}+a_2f^{-2}\,,
$$ 
with the constraint that $a_i>0$. This expression is found to be an
excellent description of the noise PSD.

In principle, the chopper synchronous offset could introduce 
correlations between declinations. 
In practice, these correlations are effectively removed by the 
$M-(L+T)/2$ differencing. 
In the vast majority of the data, $m_3$ is consistent with zero. 
It is also found that when the 
atmosphere is calm, $m_2$ is also small ($<5\%$) after LMT subtraction
and projection of corrupted modes.
These cross-sweep correlations are included in the noise correlation matrix, 
but are not found to not have a significant effect on the noise estimate. 
Nonetheless, approximately $\sim 20-30\%$ of the data was cut because it showed 
a significant residual correlation ($m_2>5\%$) between sweeps after 
atmospheric removal. 

Although the data used in the analysis are dominated by the instrument noise, 
atmospheric induced cross-channel correlations must be taken into account. 
Although they are small, the inclusion of the cross channel correlations 
can lead to a noise correlation matrix that is not guaranteed to be positive 
definite.
Including these correlations (and removing the negative eigenmodes 
in the S/N transformation [\ref{app:sn}]) 
does not appear to effect the resulting power spectrum, 
however, it does introduce an element of uncertainty to the analysis.
To avoid this complication, the band-powers presented in this paper are 
calculated without including cross-channel correlations.
This results in a small uncertainty in the noise estimate equal to the largest
measured fractional cross-channel correlation. 
By cutting an additional $10\%$ of the data observed to have a correlation 
exceeding 5\%, we can guarantee that the noise estimate is correct to this level.
From the distribution of channel correlations, we estimate the real error in 
the noise estimate to be much smaller. 
The power spectrum presented in this paper is insensitive to uncertainty in 
the noise estimate at the level of a few percent.

\section{B. Correlation and FFT}\label{app:fft}

The calculation of the various moments described in Appendix  
\ref{app:noise} can be done by Fast Fourier transformation. One 
slight complication is Fourier transformation assumes the input 
data to be periodic. For a strictly stationary process this is 
not a problem since the correlation usually vanishes after a sufficiently
long time lag. For piecewise stationary processes, a straightforward 
application of FFT produces false correlation for lag $> n/2$ , where $n$
is the number of elements. This can be prevented by 
zero padding. 

Suppose we want to calculate the correlation of
two real, $n$-component stochastic data vectors ${\bf x}$ and ${\bf y}$.
When ${\bf x}$ and ${\bf y}$ are stationary, averaging over
indices {\it is} ensemble average,
\begin{equation}
C_{ij}\equiv\langle x_i y_j\rangle=C(\Delta\equiv|i-j|)=
\frac{1}{2(n-\Delta)}\sum_{j=1}^{n-\Delta} (x_{j}y_{j+\Delta}+
x_{j+\Delta}y_j).\label{ave}
\end{equation}
A faster algorithm involves FFT. To prevent periodicity we 
pad the two data vectors with zeros up to $2n$, and Fourier 
transform:
\begin{align*}
f_u^{*}&=\frac{1}{2n}\sum_{j=1}^{2n}x_j\exp{\left(\frac{i\pi uj}{n}\right)},\\
g_u&=\frac{1}{2n}\sum_{k=1}^{2n}y_k\exp{\left(-\frac{i\pi uk}{n}\right)}.
\end{align*}
The inverse Fourier transform of $f^*g$ is
\begin{align*}
FFT^{-1}({\bf f^*g})& \equiv \frac{1}{(2n)^2}\sum_{jku=1}^{2n}x_jy_k
\exp{\left[-\frac{i\pi u}{n}(k-j)\right]}
\cdot\exp{\left(\frac{i\pi u \Delta}{n}\right)}\\
&=\frac{1}{(2n)^2}\sum_{jk=1}^{2n}x_jy_k\sum_{u=1}^{2n}
\exp{\left[\frac{i\pi u}{n}(j+\Delta-k)\right]}\\
&=\frac{1}{2n}\sum_{jk=1}^{2n}x_jy_k(\delta_{j+\Delta,k}+
\delta_{j+\Delta,k\pm 2n}+\delta_{j+\Delta,k\pm 4n}+\cdots)\\
&=\frac{1}{2n}\sum_{jk=1}^{2n}x_jy_k(\delta_{j+\Delta,k})=
\frac{1}{2n}\sum_{j=1}^{2n}x_jy_{j+\Delta}=
\frac{1}{2n}\sum_{j=1}^{n-\Delta}x_jy_{j+\Delta}.
\end{align*}
The last step uses the fact that $y_k=0$ for $k>n$. Comparing this with 
eq.~[\ref{ave}], we obtain
\begin{equation}
C(\Delta)=\frac{n}{n-\Delta}FFT^{-1}({\bf f^*g}+{\bf fg^*})
=\frac{2n}{n-\Delta} FFT^{-1}[\,{\Re}({\bf f^*g})\,].\label{fft}
\end{equation}
Note that with the correction factor $2n/(n-\Delta)$,
eq.~[\ref{fft}] is exact for $0\le\Delta\le n-1$. $C(\Delta)$
were then averaged for 20 minutes to improve signal-to-noise.

In the presence of a data cut, for example the guiding quasar, 
a similar ``pad-and-correct'' procedure can be used, such that the 
data corrupted by the quasar will be ignored in the calculation of 
correlation function.

\section{C. Theory Matrix for Non-uniform Beams}\label{app:theorymatrix}

For a given 2-D power spectrum density $P({\bf k})$, the Fourier 
transform of the temperature map $T({\hat {\bf r}})$, or the {\it spectrum} is 
$$
S({\bf k})=\sqrt{\frac{P({\bf k})}{2}}{\cal B}({\bf k})\zeta({\bf k}),
$$
where $\cal B$ is the Fourier transform of the response function at 
${\hat {\bf r}}=(x,y)$, and $\zeta$ a Gaussian random variable: 
$\zeta({\bf k})=\zeta_r({\bf k})+i\zeta_i({\bf k})$, $\langle\zeta_r^2\rangle=
\langle\zeta_i^2\rangle=1$. The realization condition requires that 
$$
P({\bf k})=P(-{\bf k});\;\;\;\;\zeta^{*}({\bf k})=\zeta(-{\bf k}).
$$
Note that $\langle\zeta^{*}({\bf k})\zeta({\bf k'})\rangle=
2\delta({\bf k}-{\bf k'})$,
which is a natural consequence of Gaussian random variables.
The realization condition does not introduce new correlation because
$\langle\zeta^{*}({\bf k})\zeta(-{\bf k})\rangle=
\langle\zeta^{*}({\bf k})\zeta^{*}({\bf k})\rangle=
\langle\zeta_r^2({\bf k})\rangle-
\langle\zeta_i^2({\bf k})\rangle=0$.
 
The temperature correlation is
\begin{gather*}
C_{T\{12\}}=\langle T^{*}({\hat {\bf r}}_1)T({\hat {\bf r}}_2)\rangle
=\iint d{\bf k}d{\bf k'}
\exp\left\{-2\pi i
[-{\bf k\cdot}{\hat {\bf r}}_1+{\bf k'}{\bf \cdot}{\hat {\bf r}}_2]\right\}\langle
S^{*}_1({\bf k})S_2({\bf k'})\rangle\\
=\iint d{\bf k}d{\bf k'}
\exp\left\{-2\pi i
[-{\bf k\cdot}{ {\hat {\bf r}}}_1+{\bf k'}{\bf \cdot}{\hat {\bf r}}_2]\right\}
{\cal B}^{*}_1({\bf k}){\cal B}_2({\bf k'})
\times\frac{\sqrt{P({\bf k})P({\bf k'})}}{2}\langle 
\zeta^{*}({\bf k})\zeta({\bf k'})\rangle\\
=FFT^{-1}\left[P({\bf k}){\cal B}^{*}_1({\bf k}){\cal B}_2({\bf k})\right]
_{({\hat {\bf r}}={\hat {\bf r}}_2-{\hat {\bf r}}_1)}.\label{ct12}
\end{gather*}

What we really want to know for band-power estimation is the derivative
of $C_{T\{12\}}$ with respect to the band-power $q_B$. This is easily done
by replacing $P$ in the equation above with proper shape functions 
determined by $\chi_{B\ell}$ defined in eq.~[\ref{qb}]. For simplicity we made 
$\chi_{B\ell}$ flat in $D_\ell$. In terms of P the filter functions
acquire corrections from the transformation between $D_\ell$ and $C_\ell$
, hence are not flat in band.
We parameterize a Gaussian beam as
$$
B_0(x,y)\propto\exp(-\frac{u}{2});\;\;\;\; u\equiv 
\frac{1}{2}\left[(\frac{x}{\alpha}+\frac{y}{\beta})/\gamma\right]^2+
\frac{1}{2}\left(\frac{x}{\alpha}-\frac{y}{\beta}\right)^2.
$$
In this parametrization, if $(\alpha,\beta,\gamma)=(\sigma,\sigma,1)$, 
$B_0$ is an axisymmetric 
Gaussian function with FWHM $\sigma\sqrt{8\ln 2}$. In the wavenumber 
domain, the total experimental response function is
$$
{\cal B}({\bf k})={\tilde \Xi}({\bf k})\times\exp\left\{
-\frac{4\pi^2}{2}\left[\frac{\gamma^2(k_x\alpha+k_y\beta)^2}{2}+
\frac{(k_x\alpha-k_y\beta)^2}{2}\right]
\right\}.
$$
Here ${\tilde \Xi}({\bf k})$ is the Fourier transform of LMT switching
pattern $\Xi({\hat {\bf r}})$ defined in eq.[\ref{lmt}]. For a pair of pixels
(1,2), 
$$
{\cal B}^{*}_1({\bf k}){\cal B}_2({\bf k})=|{\tilde \Xi}({\bf k})|^2\exp\left\{
-4\pi^2\left[(\mu_1+1)\sigma_{x}^2k_x^2+(\mu_2+1)\sigma_{y}^2k_y^2+
\mu_3\sigma_x\sigma_yk_xk_y\right]
\right\},
$$
where
$$
\mu_1\equiv \frac{[(\gamma_1^2+1)\alpha_1^2+(\gamma_2^2+1)\alpha_2^2]}
{4\sigma_x^2}-1,
$$
$$
\mu_2\equiv \frac{[(\gamma_1^2+1)\beta_1^2+(\gamma_2^2+1)\beta_2^2]}
{4\sigma_y^2}-1,
$$
$$
\mu_3\equiv \frac{[(\gamma_1^2-1)\alpha_1\beta_1+
(\gamma_2^2-1)\alpha_2\beta_2]}{2\sigma_x\sigma_y}.
$$
Here $(\sigma_x,\sigma_y)$ are taken to be the average 
of beam RMS in RA and DEC. To the zeroth order in ${\boldsymbol \mu}$, 
$$
C_{T\{12\}}({\boldsymbol{\mu}})\sim C_{T\{12\}}(0)=
FFT^{-1}\left\{P({\bf k})|{\tilde \Xi}({\bf k})|^2
\exp\left[-4\pi^2(\sigma_{x}^2k_x^2+\sigma_{y}^2k_y^2)\right]\right\}
_{({\hat {\bf r}}={\hat {\bf r}}_2-{\hat {\bf r}}_1)}.
$$
Since $\ell=4\pi \sqrt{k_x^2+k_y^2}$, if $\sigma_x=\sigma_y$, 
${\cal B}^2$ reduces to the well known beam smearing function 
$B_{\ell}=\exp{(-\sigma^2\ell^2)}$ for a symmetric Gaussian beam. 

We can obtain next order approximation of $C_{T\{12\}}$ by taking 
its partial derivatives in ${{\boldsymbol{\mu}}}$:
\begin{align*}
\frac{\partial C_{T\{12\}}}{\partial \mu_1}{\Bigg\vert}_{{{\boldsymbol{\mu}}}
=0}&=FFT^{-1}\left\{(-4\pi^2\sigma_x^2k_x^2)P({\bf k})
|{\tilde \Xi}({\bf k})|^2
\exp\left[
-4\pi^2(\sigma_{x}^2k_x^2+\sigma_{y}^2k_y^2)\right]\right\}
_{({\hat {\bf r}}={\hat {\bf r}}_2-{\hat {\bf r}}_1)};\\
\frac{\partial C_{T\{12\}}}{\partial \mu_2}{\Bigg\vert}_{{{\boldsymbol{\mu}}}
=0}&=FFT^{-1}\left\{(-4\pi^2\sigma_y^2k_y^2)P({\bf k})
|{\tilde \Xi}({\bf k})|^2\exp\left[
-4\pi^2(\sigma_x^2k_x^2+\sigma_y^2k_y^2)\right]
\right\}
_{({\hat {\bf r}}={\hat {\bf r}}_2-{\hat {\bf r}}_1)};\\
\frac{\partial C_{T\{12\}}}{\partial \mu_3}{\Bigg\vert}_{{{\boldsymbol{\mu}}}
=0}&=FFT^{-1}\left\{(-4\pi^2\sigma_x\sigma_yk_xk_y)P({\bf k})
|{\tilde \Xi}({\bf k})|^2\exp\left[
-4\pi^2(\sigma_x^2k_x^2+\sigma_y^2k_y^2)\right]
\right\}
_{({\hat {\bf r}}={\hat {\bf r}}_2-{\hat {\bf r}}_1)};\\
&\;\;\;\vdots\\
&{\rm etc.}\\
\end{align*}
The first few terms of $C_{T\{12\}}$ are 
\begin{gather*}
C_{T\{12\}}({\boldsymbol \mu})=C_{T\{12\}}({\boldsymbol \mu}=0)+\sum_{i}
\left(
\frac{\partial C_{T\{12\}}}{\partial \mu_i}{\Bigg\vert}_{{{\boldsymbol{\mu}}}=
0}\right)
\mu_i\\
+\frac{1}{2!}\sum_{ij}\left(
\frac{\partial^2 C_{T\{12\}}}{\partial \mu_i\partial \mu_j}
{\Bigg\vert}_{{{\boldsymbol{\mu}}}=0}\right)
\mu_i\mu_j+
\cdots
\end{gather*}
The largest corrections occur at high $\ell$. For $\ell\sim 2000$, FWHM
$\sim 5.5'$ and a $10\%$ beam distortion ($\mu=0.1$), the leading order 
correction is on the order of $4\pi^2\sigma^2k^2\mu\sim 20\%$ for
that specific pixel pair. However, since $(\sigma_x,\sigma_y)$ are the noise 
weighted average beam sizes, the correction goes both ways and the resulting 
power spectrum does not change significantly.

\section{D. Signal-to-Noise Eigenmode Truncation and Foreground 
Removal}\label{app:sn}

\begin{figure*}[t]
\centerline{
\psfig{figure=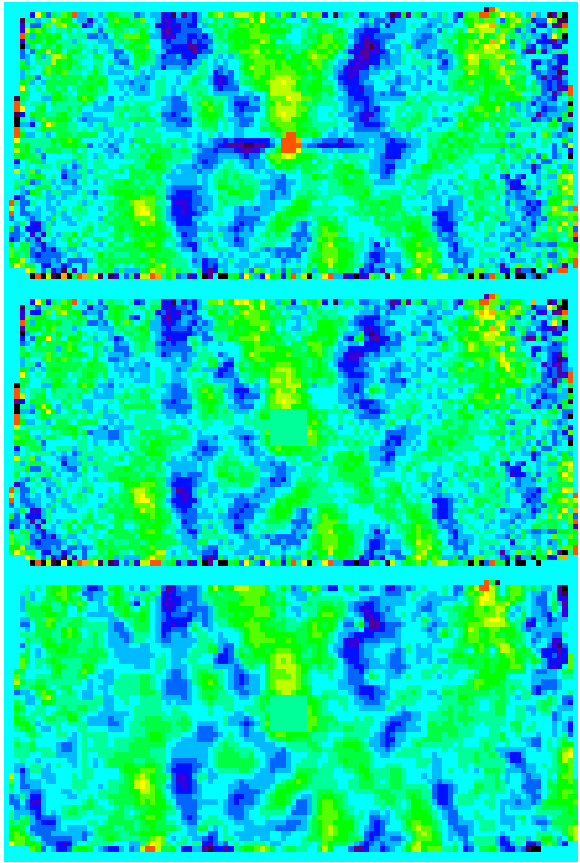,width=3in,height=4.5in,angle=0}}
\caption{
The three panels show the noise weighted coadded map  ${\bf T}$ (top), 
the foreground removed map ${ {\bf T}}'_r$ (middle), and the 
high signal-to-noise map ${ {\bf T}}'_s$ (bottom).
The top panel is the noise weighted coadded map for the CMB5 field;
it shows the bright quasar and the artifact generated from mode removal 
(the blue horizontal stripes adjacent to the quasar).
The second panel shows the same map after the foreground modes defined in 
Table \ref{tab:foreground2} are set to zero and ignored in the power spectrum
calculation. Other than the bright guiding quasar, there are no other clear 
sources of significant foreground emission. This qualitative statement is 
supported by the fact that after the removal of the guide quasar, the power 
spectrum derived from this field is unchanged by the removal of the radio
source and galactic dust templates. The bottom panel 
 shows foreground removed map constructed from only the high S/N eigenmodes.
Clearly, the S/N eigenmode transformation preserves the basic features
of the map.
}\label{fig:cmb5_raw}
\end{figure*}

The constraint matrix formalism is typically used to remove undesired 
foreground modes \citep{bond98, halverson02, myers02}. 
Combined with signal-to-noise eigenmode analysis 
(Karhunen-Lo\a' eve transformation) \citep{bunn97,bond98}, the foreground 
removal 
can be implemented in an alternative algorithm.
First we construct the foreground mode projection 
matrices that remove the modes described in Tables~\ref{tab:foreground1} 
and \ref{tab:foreground2}. 
Similar to \S\ref{subsubsec:nwcm}, if ${\bf U}$ consists of $m$ columns of 
linearly 
independent foreground modes, in our pixel space ${\bf U}$ manifests 
itself as ${\tilde {\bf U}}={\bf KU}$ (eq.~[\ref{kcorrect}]). 
Following \citet{tegmark97b}, we define the projection 
matrix ${\bf P}\equiv {\bf I}-{\tilde {\bf U}}({\tilde {\bf U}}^t
{\tilde {\bf U}})^{-1}{\tilde {\bf U}}^t$. 

Next we solve for the eigenvalues $\omega_i$ and eigenvectors ${\bf e}_i$ for
the foreground removed noise correlation 
${{\bf C}^{\prime}}_N\equiv {\bf P}{{\bf C}}_N{\bf P}$. 
The projection of ${\bf P}$ creates a $m$ - dimensional degenerate vector 
space with eigenvalue zero. Numerically, the eigenvalues of the undesired 
modes are at least 10 orders of magnitude smaller than that 
of the desired modes. 
Assume $\omega_1=\omega_2=\dots=\omega_m=0$.
The ``whitening" matrix is defined as
$$
{\bf W}\equiv
\begin{bmatrix}
\;\;\;\;\arrowvert&\;\;\;\;\arrowvert&\;&\;\;\;\;\;\;\;\arrowvert
\\
\omega_{m+1}^{-1/2}{\bf e}_{m+1}&\omega_{m+2}^{-1/2}{\bf e}_{m+2}
&\dots&\omega_n^{-1/2}{\bf e}_n
\\
\;\;\;\;\downarrow&\;\;\;\;\downarrow&\;&\;\;\;\;\;\;\;\downarrow
\end{bmatrix},
$$ 
such that ${\bf W}^t{{\bf C}}^{\prime}_N{\bf W}={\bf I}$, 
where ${\bf I}$ is the $n-m\equiv m_1$ dimensional identity matrix.
After the transformation of the temperature map by the application of
${\bf W}$, the foreground modes are 
``projected out'' completely, since eigenvectors 
(${\bf e}_1, {\bf e}_2, \cdots {\bf e}_m$) with zero eigenvalues
are not included in the construction of ${\bf W}$.
To proceed with the signal-to-noise eigenmode transformation,
we construct a fiducial theory matrix that is compared with the noise matrix to 
decide which modes contain useful information (high S/N). 
The theory matrix ${{\bf C}}_T=
\sum_B q_B {\partial { {\bf C}}_T}/{\partial q_B}$, 
where the $q_B$'s are the band powers we wish to estimate.
The most reasonable choice of a fiducial power spectrum 
is a constant flat band-power $\Sigma^2=10^{4}\,\mu{\rm K}^2$,
chosen as a conservative upper bound for $D_\ell$. 
The application of {\bf W} to the fiducial theory matrix and temperature map 
removes the corrupted foreground modes:
${{\bf C}}_T\rightarrow {\bf W}^t{{\bf C}}_T{\bf W};$ 
$\;\;\;{ {\bf T}}\rightarrow {\bf T}_r\equiv
{\bf W}^t{ {\bf T}}$. 

By operating on ${\bf W}^t{{\bf C}}_T{\bf W}$ with its orthonormal matrix 
${\bf R}$, it can be diagonalized;
$$
{\bf W}^t{{\bf C}}_T{\bf W}\rightarrow 
{\bf R}^t{\bf W}^t{{\bf C}}_T{\bf W}{\bf R}\equiv {\pmb{\mathcal E}}=
{\rm diag}({\cal E}_k).
$$
In this new basis, the theory matrix consists of diagonal elements of 
signal-to-noise values ${\cal E}_k$. 
Modes corresponding to small ${\cal E}_k$ do not significantly contribute 
to the likelihood function and only the $n_s$ high S/N 
modes are retained.
For the CMB2 field the number of modes was decreased from 9000 to 2500,
and for CMB5 the reduction was from 5000 to 2000.
Eliminating these ``noisy'' modes greatly reduced the computational resources 
required by the quadratic iteration described in 
\S\ref{subsubsec:iterative}. 
The first $n_s$ columns of 
${\bf R}$ form a $m_1\times n_s$ transformation matrix ${\bf R}_1$.
The foreground subtracted, high S/N mode data are ${\bf T}_s\equiv{\bf R}_1^t{\bf W}^t
{ {\bf T}}$; and the noise correlation matrix is the
$n_s$ dimensional identity matrix.
The resulting CMB power spectrum is unchanged for values of the cut-off 
S/N ranging from ${\cal E}_c= 0.1$ to $0.01$; this is a direct consequence 
of the conservative choice of $\Sigma^2$.
The results presented in this paper were calculated using ${\cal E}_c=0.05$.

To demonstrate the foreground removal and 
Karhunen-Lo\a' eve transformation in effect, it is useful to 
reconstruct the pixel-space map from the lower dimensional ${\bf T}_r$ 
and ${\bf T}_s$ data vectors,
\begin{align*}
{ {\bf T}}'_r&={\bf W}{\bf \Theta}{\bf T}_r;\\
{ {\bf T}}'_s&={\bf W}{\bf \Theta}{\bf R}_1{\bf T}_s,\label{maps}
\end{align*}
where ${\bf \Theta}={\rm diag}(\omega_k),\;\;\; (k=m+1,\;,m+2,\dots,n)$. 
${ {\bf T}'}_r$ is the foreground-removed map, and 
${ {\bf T}'}_s$ is the foreground-removed map reconstructed only from 
high signal-to-noise modes. 
${ {\bf T}'}_r$ and ${ {\bf T}'}_s$
are both $n$ dimensional, can be directly compared with the raw coadded 
map ${ {\bf T}}$.
The foreground removed maps for the CMB2 and CMB5 fields are shown in 
Figures~\ref{fig:cmb2_map} and \ref{fig:cmb5_map}.
For comparison, the raw coadded map, the foreground-free map, and the high 
S/N eigenmode maps for the CMB5 field are shown in Figure~\ref{fig:cmb5_raw}.


\bibliographystyle{apj}
\bibliography{merged}

\end{document}